\documentclass[journal]{IEEEtran}
\usepackage[utf8]{inputenc}
\usepackage[T1]{fontenc}
\usepackage{graphicx}
\usepackage{subfig}
\usepackage{float}
\usepackage{amsmath,amssymb,amsfonts}
\usepackage{algpseudocode,algorithm,algorithmicx,float}

\usepackage{tabularx,booktabs}
\usepackage{multirow}
\usepackage{makecell}
\setcellgapes{3pt}

\usepackage{lipsum}
\usepackage{color, xcolor}

\newcommand{\xmark}{\ding{55}}
\newcommand{\cmark}{\ding{51}}%

\usepackage{verbatim}
\usepackage[final]{pdfpages}

\usepackage{caption}
\usepackage[noadjust]{cite}
\usepackage{pifont}
\usepackage{enumitem}
\usepackage{csquotes}
\usepackage{xspace}
\usepackage{hyperref}
\usepackage{balance}
\usepackage{ulem}

\newcommand{\FedAdapt}{\texttt{FedAdapt}\xspace}

\newcommand\hl[1]{{\color{black}#1}}
\newcommand\di[1]{{\color{black}#1}}

\begin{document}

\bstctlcite{IEEEexample:BSTcontfrol}

\title{\FedAdapt: Adaptive Offloading for IoT Devices in Federated Learning}

\author{
	    Di Wu,
	    Rehmat Ullah,
	    Paul Harvey,
	    Peter Kilpatrick,
	    Ivor Spence,
	    and Blesson Varghese

	    \thanks{
	    %Copyright (c) 20xx IEEE. Personal use of this material is permitted. However, permission to use this material for any other purposes must be obtained from the IEEE by sending a request to pubs-permissions@ieee.org.

	    This work was sponsored by funds from Rakuten Mobile, Japan. The last author was also supported by a Royal Society Short Industry Fellowship.

	    D. Wu, R. Ullah and B. Varghese are with the School of Computer Science, University of St Andrews, UK.

	    P. Kilpatrick and I. Spence are with the School of Electronics, Electrical Engineering and Computer Science, Queen’s University Belfast, UK.

	    P. Harvey is with Autonomous Networking Research \& Innovation Department, Rakuten Mobile, Japan.
	    }
}

\maketitle
\begin{abstract}
Applying Federated Learning (FL) on Internet-of-Things devices is necessitated by the large volumes of data they produce and growing concerns of data privacy. However, there are three challenges that need to be addressed to make FL efficient: (i) \hl{execution} on devices with limited computational capabilities, (ii) \hl{accounting} for stragglers due to computational heterogeneity of devices, and (iii) \hl{adaptation} to the changing network bandwidths. This paper presents \FedAdapt, an adaptive offloading FL framework to mitigate the aforementioned challenges. \FedAdapt accelerates local training in computationally constrained devices by leveraging layer offloading of deep neural networks (DNNs) to servers. Further, \FedAdapt adopts reinforcement learning based optimization and clustering to adaptively identify which layers of the DNN should be offloaded for each individual device on to a server to tackle the challenges of computational heterogeneity and changing network bandwidth.
%Experimental studies are carried out on a lab-based testbed comprising five IoT devices.
Experimental studies are carried out on a lab-based testbed and it is demonstrated that by offloading a DNN from the device to the server \FedAdapt reduces the training time of a typical IoT device by over half compared to classic FL. The training time of extreme stragglers and the overall training time can be reduced by up to 57\%.
Furthermore, with changing network bandwidth, \FedAdapt is demonstrated to reduce the training time by up to 40\% when compared to classic FL, without sacrificing accuracy.
\end{abstract}


\section{Introduction}
\label{sec:introduction}
Internet of Things (IoT) devices generate large volumes of data from user devices that \hl{are} often deemed sensitive. 
For example, wearable devices such as the Google Glass or Apple watch gather sensitive data by recording the daily activities of users~\cite{perera2015big}.
This data can be analyzed using machine learning (ML) techniques for delivering personalized services~\cite{zhou2019edge,qian2020orchestrating}.
Privacy preserving ML techniques are required to ensure that sensitive data can be analyzed in a safe manner. 

Federated Learning (FL) is a privacy preserving ML technique that has recently gained popularity~\cite{yang2019federated,kairouz2019advances,briggs2020review}. Using this technique, an ML model, for example, a Deep Neural Network (DNN) is executed on several IoT devices. The model on each device is trained without sending raw data (that may be sensitive) from the device to a server. Instead, the server receives intermediate models generated by the devices that are aggregated on the server to create a global model. Thus, an ML model can be trained by not exposing sensitive data from a device to an external server. 

In the classic FL architecture, the computationally intensive workload (training of the DNN) is executed on the device. A server located at the edge of the network or on the cloud only aggregates the weights sent from the devices, which is relatively less computationally intensive. Aggregation is required for updating a global model on the server that is then sent to the devices to continue training.
The devices train independently and may be connected to the server through various network configurations. 

Classic FL is however limited in the following three ways:

(1) \textit{Impractical training times on computationally constrained IoT devices}: 
The computationally intensive workload of training \hl{used in} FL is required to be executed on devices that are relatively resource constrained when compared to large servers or clusters that may have specialized processors for training ML models. Thus, the time taken to train ML models on devices can be large making FL impractical for real-world scenarios~\cite{das2019privacy,wang2020towards,gao2020end}. For example, a lightweight convolutional neural network, MobileNetV1~\cite{Howard2017MobileNetsEC} required over 8 hours on Raspberry Pi3 single board computers to complete one round of training in FL~\cite{gao2020end}. 
Therefore, there is an immediate need for techniques that accelerate training in FL on IoT devices. 

(2) \textit{Stragglers arising from computational heterogeneity of IoT devices can slow down other devices during training}:
IoT devices connected to a server for FL may have varying computational capabilities or heterogeneous architectures. Devices that require a longer time for training, referred to as stragglers in this article, will slow down all devices in a synchronous FL system~\cite{li2019smartpc,dhakal2019coded,chen2019asynchronous}. This is because the aggregating server will need to wait until all devices have completed training~\cite{imteaj2020federated}. Asynchronous FL systems have been developed to mitigate the straggler problem. However, they affect the accuracy of the models since all devices may not contribute equally to training~\cite{xu2019helios,hakimi2019taming}. Therefore, approaches that reduce the impact of stragglers are required. 

(3) \textit{Varying operational conditions can increase training time}:
Operational conditions, such as the network bandwidth between a device and the server can vary during the course of training. This can impact the training time~\cite{bonawitz2019towards,li2019smartpc}. These need to be considered for efficient FL training and therefore adaptive context-aware strategies that account for changing operational conditions are \hl{needed}. 

The research presented in this paper aims to address the above challenges by considering the following questions:
\begin{itemize}
    \item \textbf{RQ1}: What techniques can be adopted to accelerate FL in an IoT environment?
    \item \textbf{RQ2}: What techniques can be adopted to minimize the impact of computational heterogeneity of IoT devices?
    \item \textbf{RQ3}: How can \hl{these} techniques adapt to changes in network conditions?
\end{itemize}

This paper presents \FedAdapt, a holistic framework that mitigates the above challenges of accelerating FL, reducing the impact of \hl{computational heterogeneity} and adapting to varying network bandwidth. 
To accelerate FL training and to address \textbf{RQ1}, \FedAdapt is underpinned by an offloading technique in which the layers of a DNN model can be offloaded from a device to a server to alleviate the computational burden of training on the device. \hl{To address \textbf{RQ2}}, \FedAdapt incorporates a Reinforcement Learning (RL) strategy to automate the identification of layers that are offloaded from a device to the server. To address \textbf{RQ3}, \FedAdapt further optimizes the RL strategy to develop different offloading strategies for each device while accounting for changing network bandwidth. A clustering technique is used to rapidly generate the offloading strategy.

The above are developed on a testbed comprising five devices and a server for two DNN models, namely VGG-5 and VGG-8~\cite{simonyan2014very}. 
The experimental studies highlight that \FedAdapt introduces a negligible overhead (time for executing \FedAdapt modules).
The key results are that \FedAdapt reduces the total training time of VGG-5 by 30\% while achieving the same accuracy and convergence speed when compared to classic FL. \hl{Additionally, \FedAdapt} reduces the total training time of VGG-8 by 40\% when compared to FL without requiring further RL training. 

The \textbf{\textit{research contributions}} of this paper are:
%\begin{itemize}
%  \item  The development of a DNN layer offloading strategy in FL to accelerate training. Over 45\% reduction in training time is observed by using the offloading strategy on a typical IoT device. 
%  \item The development of an adaptive technique using RL that generates optimal offloading strategy for heterogeneous devices to reduce the impact of stragglers. The training time on the experimental test bed was observed to reduce by 40\% for each round of FL using VGG5. An overhead of 0.5\%, which is the time for executing the \FedAdapt modules is noted, but is negligible compared to the gain. A performance gain of 57\% was observed when using \FedAdapt for VGG8. 
%  \item The development of \FedAdapt, a holistic framework that incorporates techniques for accelerating FL training, reducing the impact of stragglers, and adapting to varying network bandwidth. \FedAdapt outperforms classic FL by reducing the training time by up to 40\%.  
%\end{itemize}

%(1) The \textit{development of a DNN layer offloading strategy} in FL to accelerate training. Over 45\% reduction in training time is observed using the offloading strategy on an IoT device.}

(1) The \textit{development of an adaptive offloading technique} that generates optimal offloading strategies \hl{for devices to reduce the impact of computational heterogeneity}. The training time on the experimental test bed was observed to reduce by 40\% for each round of FL using VGG-5.
%An overhead of 0.5\%, which is the time for executing the \FedAdapt modules is noted, but is negligible compared to the gain. 
\hl{This is achieved by the first ever introduction of a variable layer-wise training strategy to FL, which relies on offloading by using RL.} To the best of our knowledge, \FedAdapt is the first work to introduce dynamic offloading strategies into FL training.

(2) The \textit{development of \FedAdapt}, a holistic framework that incorporates techniques for accelerating FL training, reducing the \hl{impact of computational heterogeneity,} and adapting to varying network bandwidth. \FedAdapt outperforms classic FL by reducing the training time by up to 40\%.  

The rest of this article is organized as follows. Section~\ref{sec:background} presents the background and related work. Section~\ref{sec:fedadapt} presents the \FedAdapt framework and the problem model. 
Section~\ref{sec:fedadapt-rl} presents the training process of a reinforcement learning agent essential to \FedAdapt.
Section~\ref{sec:evaluation} highlights the results obtained from an experimental study. 
Section~\ref{sec:conclustion} concludes this paper and presents future work. 

\section{Background And Related Work}
\label{sec:background}
\begin{figure}
		\centering
		\includegraphics[width=0.44\textwidth]{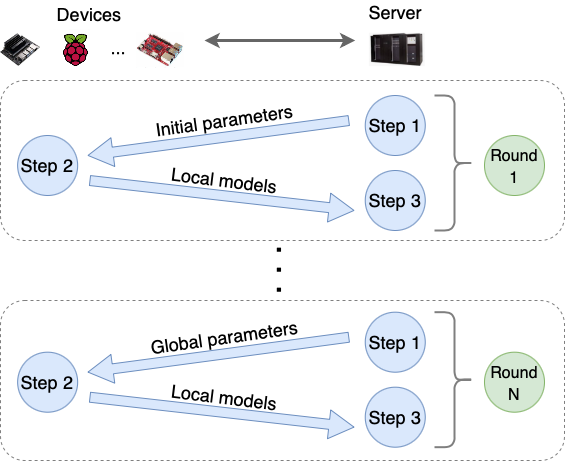}
		\caption{Steps in each round of FL training: Step 1 - Server initializes the parameters of the global model and sends to each device, Step 2 - Each device completes training on its local dataset and sends local model to server, and Step 3 - Server aggregates local models to generate a new global model.}
		\label{fig1:flsteps}
\end{figure}

\textit{Federated Learning: } 
FL~\cite{mcmahan2017communication} was developed to train machine learning (ML) models in a distributed manner. 
Each round of FL comprises three key steps as shown in Figure~\ref{fig1:flsteps}.
In the first step, a global model is initialized on the server and distributed to all devices. Each device independently trains the ML models using data generated by the device. Typically, one epoch of local training utilizes the entire dataset from each device. 
After independently training, in the second step, the trained models from the local devices (updated model parameters) are sent to the server.
In the third step, a new global model is aggregated using methods such as Federated Averaging (FedAvg) on the server~\cite{mcmahan2017communication}. 
In subsequent rounds of FL training, the aggregated model is distributed to all devices and the above steps are repeated until the training loss converges or a time limit is exceeded. FL is scalable on each device since local training can be carried out independently. However, FL is known to be less efficient for heterogeneous devices that have different computational capabilities. Thus local training in the straggler becomes a bottleneck~\cite{gao2020end,wang2020towards}.

\textit{Split Learning: } 
SL~\cite{vepakomma2018split} was developed to partition a monolithic DNN into two networks, namely a device-side and a server-side network. 
On the device-side, the DNN is trained up to the layer at which the DNN is partitioned. \hl{Then the activation feature map of the last layer on the device is sent to the server.} The server continues training until the last layer of the DNN. After the training loss is calculated and the gradient is updated, the respective gradients are sent to the device \hl{so that the gradients on the device-side can be calculated and updated.}
When training in SL with multiple devices, the devices are trained in a sequential round robin fashion whereby only one device will be connected to the server at a time. After a given device completes training, the updated weights are copied onto the next device to continue training. By training with fewer layers on the device-side, computation can be significantly reduced on the device compared to FL in which the entire DNN is trained on the device. 
While SL is beneficial for collaborative training when there are a small number of devices, it is inefficient for a large number of participating devices due to sequential training across the devices. 

\textit{\hl{Splitfed Learning: }} 
The synergy of FL and SL \hl{has} been explored recently to mitigate the above limitations. \hl{Splitfed Learning (SFL)} was proposed to achieve both parallel training of FL and acceleration of device training in SL~\cite{thapa2020splitfed}. 
The combination of SFL and transfer learning has been proposed to further improve the convergence rate of large models (e.g. ResNet56) on limited resources~\cite{he2020group}. A local loss is incorporated in the device side to reduce the communication overhead~\cite{hanaccelerating}. However, existing SFL-based research does not consider optimal partitioning strategies or require hardware configuration data to manually determine the model partitions for all devices before training. In addition, a static partition strategy could become sub-optimal when operational conditions change during training.

\textit{Computation offloading in edge computing:}
Computation offloading has been widely adopted in the literature of edge computing. Computationally expensive components of a distributed application that need to be executed on resource constrained devices are offloaded to a server located nearby thereby alleviating the computational burden on the device~\cite{zheng2020survey}.
Research on optimizing the offloading strategy to maximize performance (e.g. training time) while minimizing energy consumption has been explored~\cite{ali2019deep}. However, the application of computation offloading to machine learning tasks on edge devices is still in the initial stages of exploration, particularly within the context of FL training. There is research on determining optimal offloading for inference~\cite{scission,kang2017neurosurgeon}. However, since FL training is computationally intensive and requires more time than an inference query, a more adaptable and dynamic offloading strategy that reacts to changes in operational conditions is required.

%It is observed that applying FL on resource constrained and heterogeneous devices has been minimally explored. The end-to-end performance of FL on homogeneous single board computers has been explored~\cite{gao2020end}. There is recent research on FL optimization for heterogeneity. The ELFISH framework adjusts the size of DNN models based on the computational resources on each device~\cite{xu2019helios}. A data-aware technique is proposed to schedule different DNN workloads on heterogeneous mobile devices~\cite{wang2020towards}. However, the offloading strategy is based on a prior knowledge of the training time that is estimated. This is limited in real deployments since the operational conditions of the devices may change over time.

\begin{table} 
\begin{center}
 \caption{Comparison of FL, SL, SFL and \FedAdapt}
\begin{tabular}{*{5}{c}}
\hline
 & FL & SL & SFL &\FedAdapt \\
\hline\hline
Independent (parallel) training  & \cmark & \xmark & \cmark & \cmark\\
Limited computational resources  & \xmark & \cmark & \cmark & \cmark\\
Heterogeneous devices  & \xmark & \xmark & \xmark & \cmark \\
Changing network bandwidth  & \xmark & \xmark  & \xmark & \cmark\\
Optimizing offloading strategy  & \xmark & \xmark  & \xmark & \cmark\\ 
\hline
\end{tabular}
\label{works}
\end{center}
\end{table}

\textit{How does \FedAdapt differ from prior work?} 
%FL, SL and \FedAdapt are inherently privacy preserving. 
%We note from the literature that most FL, SL and SFL implementations are simulation-based and do not focus on real test beds. \FedAdapt on the other hand is developed and demonstrated on a physical lab-based test bed. 
Table~\ref{works} presents a comparison of FL, SL, SFL and \FedAdapt.
As in FL, \FedAdapt independently trains on the local device and\hl{, as in SL,} accounts for the limited computational resources on the device. 
Although DNN layer offloading is also utilized in SFL, the key differences are that \FedAdapt accounts for heterogeneous devices and \hl{are changing} network bandwidth that affect training performance not considered within classic SFL. 
%While \FedAdapt leverages the advantages of FL and SL, DNN layer offloading is utilized to accelerate local training on resource limited devices. 
In addition, \FedAdapt requires no prior knowledge of the devices, but uses an automated approach based on RL to identify the DNN partitions for each device, thus mitigating the challenge of heterogeneity. 
We also note from the literature that most FL, SL and SFL implementations are simulation-based and do not focus on real test beds. The benefits of \FedAdapt on the other hand are demonstrated on a physical lab-based test bed.

\section{The FedAdapt Framework}
\label{sec:fedadapt}
This section provides an overview of the \FedAdapt framework and the underpinning techniques. Then the problem of distributing the DNN model in FL across the device and server for performance efficiency is formulated. 

\subsection{Overview}
\label{subsec:overview}

%\FedAdapt is a holistic framework for adaptively offloading partial training workloads in FL from straggler IoT devices on to server resources located at the edge of the network. Thus, not only is local training in FL rounds accelerated but the framework automates the approach for reducing the impact of stragglers during each FL round and for adapting to changing network bandwidth between the devices and the server.

%Consider a computer vision-based use-case in which a global model of landmarks\footnote{\url{https://cloud.google.com/vision/docs/detecting-landmarks}} in a city needs to be trained for detecting streets and man-made structures. A comprehensive model can be developed by capturing data from multiple disparate sources, including street cameras, self-driving cars, smart wearable glasses and smartphones. Here FL techniques can be employed to preserve data privacy while collaboratively training the model from different data sources. \FedAdapt can be employed in this use-case given that some devices may be relatively more computationally constrained than others (and can slow down the training process) and will have varying network bandwidth between the devices and the server.}

\begin{figure}
		\centering
		\includegraphics[width=0.44\textwidth]{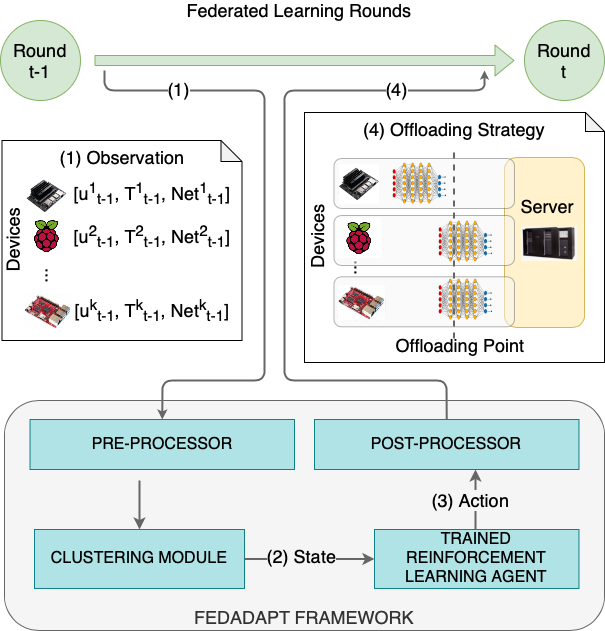}
		\caption{\FedAdapt framework and its positioning within FL.}
		\label{fig2:overview_FedAdapt}
\end{figure}

\FedAdapt (Figure~\ref{fig2:overview_FedAdapt}) comprises four modules, namely (1) Pre-processor, (2) Clustering Module, (3) Trained Reinforcement Learning (RL) Agent, and (4) Post-processor. 

After a FL round has been completed (Round $t-1$), the \textit{Pre-processor} gathers observation on the state of the devices, such as computational capabilities\footnote{In this article, we define computational capabilities as the training time per iteration for one batch of training samples} and network bandwidth between each device and server. The training time per iteration is normalized by the Pre-processor. 

The \textit{Clustering Module} groups devices with similar training time into a single group. In other words, all devices within a group are considered as computationally homogeneous. A group is further defined by accounting for network bandwidth between the device and the server.
RL is employed to determine the offloading strategy for each group.

The \textit{Trained RL Agent} given the group information and observations (referred to as State) will generate an offloading decision (referred to as Action) for each group by using a fully-connected neural network. The training process of the RL Agent is further discussed in Section~\ref{sec:fedadapt-rl}.

The \textit{Post-processor} makes use of the output of the Trained RL Agent and maps the offloading decision for each group on to the devices in the group. All devices in a group execute the same offloading strategy. The offloading strategy indicates which layers of the DNN model will be on each device for the FL Round $t$.

\FedAdapt is underpinned by three techniques. The first technique is \textit{offloading} in which the DNN model used in FL is partitioned so that certain layers of a neural network can be offloaded from computationally constrained devices on to the server. The offloading approach is used to accelerate FL training since the computational workload is transferred to more capable resources that may be available on the server; \hl{this} addresses RQ1 posed in Section~\ref{sec:introduction}. The performance gain obtained by offloading will be experimentally shown in the next section. 

In the offloading approach, the layer after which the DNN model is partitioned is referred to as the Offloading Point (OP). The OP is identified by the RL Agent. The initial layers of the DNN remain on the device whereas the layers after the OP are offloaded to the server. During training, the intermediate activation and corresponding labels and gradients of the distributed DNN are exchanged between the devices and server. Although there are communication overheads in transferring the activation and gradients during training, the overall FL training time is reduced due to the gain by computational offloading (refer Section~\ref{sec:evaluation}). 

% The second technique uses \textit{Reinforcement Learning} (RL) to address the challenge of computational heterogeneity of devices that leads to stragglers in FL as posed in RQ2.
The second technique uses \textit{RL} to address the challenge of computational heterogeneity of devices that leads to stragglers in FL as posed in RQ2.
This is an automated technique that enables the Post-Processor to identify the OP for each individual device before a round of FL so that an optimized offloading strategy is executed for each device participating in an FL training round.
To address the challenges in scaling training for a large number of devices and in determining an offloading strategy for all devices, a \textit{clustering-based approach} is employed to group devices that have similar computational performance.
Once the offloading decision for each group is determined, the Post-Processor maps the decision on to each device so that the offloading strategy is executed for FL training.

The third technique employed in \FedAdapt is \textit{optimized RL} so that operational conditions, namely network bandwidth between devices and the server can be accounted for generating optimal offloading strategies. Thus RQ3 raised initially is addressed by \FedAdapt. 

\subsection{Problem Model}
\label{subsec:formulation}

\FedAdapt assumes that the network bandwidth between the device and the server can change between different FL rounds. The network bandwidth from the previous FL round is observed for generating an offloading strategy. However, any changes to the network bandwidth during a round \hl{are} not accounted for. The goal is to reduce the overall training time by achieving suitable offloading strategies for all devices and adapting to network changes that are observed.

\begin{table}[t]
	\caption{Notation used in \FedAdapt}
	\centering
	\begin{tabular}{|c|c|}
		\hline
		\textbf{Notation}                       & \textbf{Description}                                                                                                        \\ \hline
		\multicolumn{1}{|m{0.8cm}|}{$K$} & 
		\multicolumn{1}{m{6.5cm}|}{Number of devices in a FL task} 
		\\ \hline
		%\multicolumn{1}{|m{0.8cm}|}{\di{\sout{$R$}}} & 
		%\multicolumn{1}{m{6.5cm}|}{\di{\sout{Total training rounds in an FL task}}} 
		%\\ \hline
		%\multicolumn{1}{|m{0.8cm}|}{$M$} & 
		%\multicolumn{1}{m{6.5cm}|}{The DNN Model used in FL} 
		%\\ \hline
		\multicolumn{1}{|m{0.8cm}|}{$t$} & 
		\multicolumn{1}{m{6.5cm}|}{Time step at round $t$} 
		\\ \hline
		\multicolumn{1}{|m{0.8cm}|}{$k$} & 
		\multicolumn{1}{m{6.5cm}|}{Denote a device participating in FL} 
		%\\ \hline
		%\multicolumn{1}{|m{0.8cm}|}{$k_{th}$} & 
		%\multicolumn{1}{m{6.5cm}|}{Index of device $k$} 
		%\\ \hline
		%\multicolumn{1}{|m{0.8cm}|}{$N^k$} & 
		%\multicolumn{1}{m{6.5cm}|}{\di{The amount of dataset in device k}}
		\\ \hline
		\multicolumn{1}{|m{0.8cm}|}{$W^k$} & 
		\multicolumn{1}{m{6.5cm}|}{Training workload of device $k$}
		\\ \hline
		\multicolumn{1}{|m{0.8cm}|}{$Net_t^k$} & 
		\multicolumn{1}{m{6.5cm}|}{Network speed at round $t$ of device $k$} 
		\\ \hline
		\multicolumn{1}{|m{0.8cm}|}{$C_t^k$} & 
		\multicolumn{1}{m{6.5cm}|}{Training speed at round $t$ of device $k$} 
		%\\ \hline
		%\multicolumn{1}{|m{0.8cm}|}{\di{\sout{$f$}}} & 
		%\multicolumn{1}{m{6.5cm}|}{\di{\sout{A function calculating training time of a device}}}
		\\ \hline
		\multicolumn{1}{|m{0.8cm}|}{$L(\mu_t^k)$} & 
		\multicolumn{1}{m{6.5cm}|}{Total amount of communication of device $k$ at round $t$}
		\\ \hline
		\multicolumn{1}{|m{0.8cm}|}{$s$} & 
		\multicolumn{1}{m{6.5cm}|}{Server coordinating the FL task} 
		\\ \hline
		\multicolumn{1}{|m{0.8cm}|}{$C_t^s$} & 
		\multicolumn{1}{m{6.5cm}|}{Training speed at round $t$ of server $s$} 
		\\ \hline
		\multicolumn{1}{|m{0.8cm}|}{$\mu_t^k$} & 
		\multicolumn{1}{m{6.5cm}|}{Offloading strategy at round $t$ of device $k$} 
		\\ \hline
		\multicolumn{1}{|m{0.8cm}|}{$\mu_t$} & 
		\multicolumn{1}{m{6.5cm}|}{Offloading strategy at round $t$ for all devices or groups}
		\\ \hline
		\multicolumn{1}{|m{0.8cm}|}{$T_t^{k}$} & 
		\multicolumn{1}{m{6.5cm}|}{Local training time at round $t$ of device $k$}
		\\ \hline
		\multicolumn{1}{|m{0.8cm}|}{$B^{k}$} & 
		\multicolumn{1}{m{6.5cm}|}{Training time of device $k$ without offloading}
		\\ \hline
		\multicolumn{1}{|m{0.8cm}|}{$T_t$} & 
		\multicolumn{1}{m{6.5cm}|}{FL training time at $t$} 
		\\ \hline
		\multicolumn{1}{|m{0.8cm}|}{$S_t$} & 
		\multicolumn{1}{m{6.5cm}|}{State at round $t$ in RL} 
		\\ \hline
		\multicolumn{1}{|m{0.8cm}|}{$A_t$} & 
		\multicolumn{1}{m{6.5cm}|}{Action generated by RL at round $t$} 
		\\ \hline
		\multicolumn{1}{|m{0.8cm}|}{$R_t$} & 
		\multicolumn{1}{m{6.5cm}|}{Reward at round $t$} 
		\\ \hline
		\multicolumn{1}{|m{0.8cm}|}{$G$} & 
		\multicolumn{1}{m{6.5cm}|}{Total number of groups}
		\\ \hline
		\multicolumn{1}{|m{0.8cm}|}{$g$} & 
		\multicolumn{1}{m{6.5cm}|}{A representative device in group $g$} 
		%\\ \hline
		%\multicolumn{1}{|m{0.8cm}|}{$g_{th}$} & 
		%\multicolumn{1}{m{6.5cm}|}{Index of group $g$} 
		\\ \hline
		\multicolumn{1}{|m{0.8cm}|}{$W^g$} & 
		\multicolumn{1}{m{6.5cm}|}{Training workload of the representative device of $g$} 
		\\ \hline
		\multicolumn{1}{|m{0.8cm}|}{$Net_t^g$} & 
		\multicolumn{1}{m{6.5cm}|}{Network bandwidth of the representative device of $g$ at round $t$} 
		\\ \hline
		\multicolumn{1}{|m{0.8cm}|}{$C_t^g$} & 
		\multicolumn{1}{m{6.5cm}|}{Training speed of the representative device of $g$ at round $t$}
		\\ \hline
		\multicolumn{1}{|m{0.8cm}|}{$\mu_t^g$} & 
		\multicolumn{1}{m{6.5cm}|}{Offloading strategy of the representative device of $g$ at round $t$} 
		\\ \hline
		\multicolumn{1}{|m{0.8cm}|}{$T_t^{g}$} & 
		\multicolumn{1}{m{6.5cm}|}{Training time of the representative device of $g$ at round $t$}
		\\ \hline
		\multicolumn{1}{|m{0.8cm}|}{$f_{norm}$} & 
		\multicolumn{1}{m{6.5cm}|}{Normalization function used in RL to calculate $R_t$}
		\\ \hline
	\end{tabular}%
	\label{table:notions}
\end{table}

\di{Assume that FL training is carried out with $K$ devices, each device has a training workload $W^k$ for each round, an FL task involving a server $s$ has training speed $C_t^s$ at round $t$, a set of participating devices $\{k\}_{k=1}^K$ have training speed $C_t^k$, and network bandwidth between the device and the server $Net_t^k$. The offloading strategy for the device is $\mu_t^k$ denoted as the remaining proportion of computation on each device at round $t$. For a round of FL training on device $k$, the proportion of the workload that is executed on the device is $\mu_t^kW^k$ and $(1-\mu_t^k)W^k$ is offloaded to the server.
Let $L(\mu_t^k)$ be the size of the feature maps that are transferred between the device and server during the training of round $t$. It is worth noting that $L(\mu_t^k)$ depends on $\mu_t^k$ as the offloading strategy determines the size of the transferred feature map. Finally, the training time for device $k$ of round $t$ may be calculated as follows:} 
\begin{equation} 
     \di{T_t^{k} = \frac{\mu_t^kW^k}{C_t^k} + \frac{(1-\mu_t^k)W^k}{C_t^s} + \frac{L(\mu_t^k)}{Net_t^k}}
     \label{eq:T_t_k}
\end{equation}
\di{where the $\frac{\mu_t^kW^k}{C_t^k}$ and $\frac{(1-\mu_t^k)W^k}{C_t^s}$ is the training time on the device and on the server, respectively. $\frac{L(\mu_t^k)}{Net_t^k}$ is the communication time during training.}

%The training time for device $k$ of round $t$ is calculated with a function $f$ as follows: 
%uation} 
     %T_t^{k} = f(W^k,C_t^s,C_t^k,Net_t^k,\mu_t^k)
     %\label{eq:T_t_k}
%\end{equation}
%where $f$ maps $W^k,C_t^s,C_t^k,Net_t^k$ and $\mu_t^k$ to training time $T_t^{k}$.

\di{In round $t$, $W^k$, $C_t^s$, $C_t^k$ and $Net_t^k$ are either \hl{constants} or are variables that are not controlled by \FedAdapt. The offloading strategy for each device $\mu_t^k$ is controlled by \FedAdapt. In this paper, $\mu_t^k$ is an OP. 
The collection of OPs for $K$ devices is $\mu_t$, which is $\{\mu_t^k\}_{k=1}^K$. In terms of SL and SFL, $\mu_t^k$ is uniform among all devices for all rounds, denoted as $\overline{\mu}$. In synchronous training, the server waits for all devices to complete training. The FL training time of round $t$ is $T_t = \max \{ \{T_t^{k}\}_{k=1}^K\}$. Table~\ref{time_complex} summarises the computational workload of $K$ devices and the training time required for one round for different methods.}

\begin{table}
\begin{center}
\caption{Computational workload (on the device) and training time for one round}
%\begin{tabular}{*{3}{l}}
\begin{tabular}{| m{1.1cm} | m{1.3cm}| m{5.15cm} |}
\hline
Methods                 &Computation                &Training Time                       \\
\hline
 FL                      &$\sum\limits_{k=1}^{K}W^k$                   &$\max\{\{\frac{W^k}{C_t^k}\}_{k=1}^K\}$                 \\
\hline
SL                      &$\sum\limits_{k=1}^{K}\overline{\mu}W^k$                   &$\sum\limits_{k=1}^{K}\frac{\overline{\mu}W^k}{C_t^k} + \frac{(1-\overline{\mu})W^k}{C_t^s} + \frac{L(\overline{\mu})}{Net_t^k}$                   \\
\hline
SFL               &$\sum\limits_{k=1}^{K}\overline{\mu}W^k$             &$\max\{\{\frac{\overline{\mu}W^k}{C_t^k} + \frac{(1-\overline{\mu})W^k}{C_t^s} + \frac{L(\overline{\mu})}{Net_t^k}\}_{k=1}^K\}$                \\
\hline
\FedAdapt                    &$\sum\limits_{k=1}^{K}\mu_t^kW^k$   &$\max\{\{\frac{\mu_t^kW^k}{C_t^k} + \frac{(1-\mu_t^k)W^k}{C_t^s} + \frac{L(\mu_t^k)}{Net_t^k}\}_{k=1}^K\}$                   \\
\hline
\end{tabular}
\label{time_complex}
\end{center}
\end{table}

%To reduce the training time for all devices in a round (different from minimizing $T_t$ in Equation~\ref{eq:max}), 
To reduce the training time for all devices in a round, we define our optimization target as minimizing the average training time for $K$ devices as follows:
\begin{equation}
\di{
\begin{aligned}
& \underset{\mu_t^k}{\text{minimize}}
& & \frac{1}{K} \sum_{k=1}^{K} T_t^{k} \\
& \text{subject to}
& & T_t^{k} = \frac{\mu_t^kW^k}{C_t^k} + \frac{(1-\mu_t^k)W^k}{C_t^s} + \frac{L(\mu_t^k)}{Net_t^k}
\label{eq:goal1}
\end{aligned}
}
\end{equation}
\di{The training time of round $t$ is $T_t$, which is bound by the maximum training time of all participating devices. However, \FedAdapt not only optimizes the maximum training time, which may be bound by the straggler devices, but also aims at reducing the training time of each device. Reducing the training time on individual devices implies reducing the amount of computation carried out on the devices. Therefore, in \FedAdapt, we define the objective as average training time for $K$ devices. The objective of reducing the total training time over all FL rounds is achieved by lowering the average training time of each round for which $\mu_t$ is optimized for each round based on variable operational conditions, namely $C_t^s$, $C_t^k$ and $Net_t^k$.}

%\begin{equation}
%\begin{aligned}
%& \underset{\mu_t^k}{\text{minimize}}
%& & \frac{1}{K} \sum_{k=1}^{K} T_t^{k} \\
%& \text{subject to}
%& & T_t^{k} = f(W^k,C_t^s,C_t^k,Net_t^k,\mu_t^k)
%\label{eq:goal1}
%\end{aligned}
%\end{equation}

%The objective of reducing the total training time over all FL rounds is achieved by lowering the average training time of all rounds for which $\mu_t$ is optimized for each round based on variable operational conditions $C_t^s$, $C_t^k$ and $Net_t^k$.
%\begin{equation}
%\begin{aligned}
%& \underset{\mu_t}{\text{minimize}}
%& & \frac{1}{R} \sum_{t=1}^{R} T_t \\
%& \text{subject to}
%& & T_t = \max \{T_t^{k}\}_{k=1}^K \\
%& & & T_t^{k} = f(W^k,C_t^s,C_t^k,Net_t^k,\mu_t^k) \\
%& & & \mu_t = \{\mu_t^k\}_{k=1}^K.
%\label{eq:goal2}
%\end{aligned}
%\end{equation}

\section{Training the Reinforcement Learning Agent for FedAdapt}
\label{sec:fedadapt-rl}
In this section, the training process of the RL Agent that is employed in \FedAdapt to achieve the objectives of Equation~\ref{eq:goal1} combined with a clustering technique is presented. For training the RL Agent, the input state, output action and the reward function, which are essential components of the RL technique are presented.

\di{RL is a sequential decision optimization technique used in a variety of domains~\cite{gu2017deep,sallab2017deep,sharma2017literature}, including optimization problems requiring automated control~\cite{mao2019park}.
An RL-based agent is used in \FedAdapt for two reasons. (1) RL provides an automated mechanism to generate reasonable offloading strategies for the participating devices, which maximizes the reward i.e., training time in \FedAdapt. Given the computational heterogeneity of IoT devices, existing research assumes that the hardware configuration of all devices can be obtained (white box), for identifying the stragglers~\cite{li2019smartpc,xu2019helios}. However, obtaining the hardware configuration of all devices may not be possible in a real FL application. In addition, the one-time estimation of training time of a device is also often inaccurate since certain factors, such as resource availability and network bandwidth may change during training. RL would eliminate the need for explicitly profiling the hardware on the device by using the training record from the last round.
(2) A trained RL Agent can be reused for similar FL tasks. We verify the performance of reusing the RL Agent without retraining in Section~\ref{subsubsec:reusingagent}. Without training for another specific model, the RL Agent can achieve 57\% training time reduction per round.}\\

\begin{figure*}
		\centering
		\includegraphics[width=0.94\textwidth]{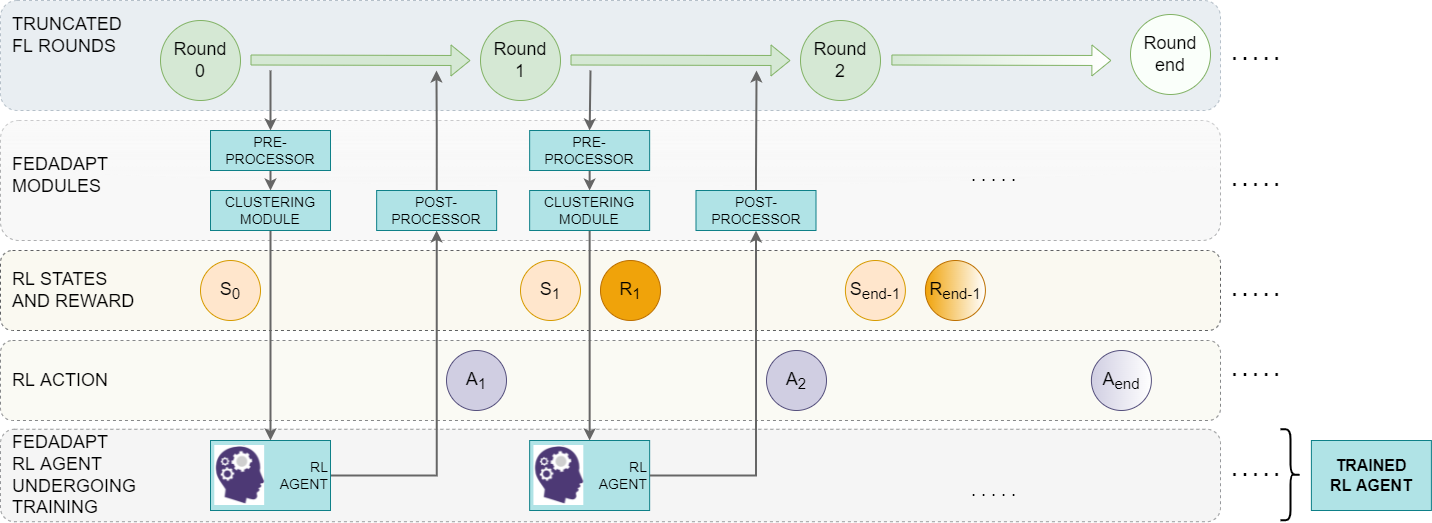}
		\caption{Training of the RL Agent used in \FedAdapt}
		\label{fig:rl}
\end{figure*}

\subsubsection*{Basic training approach}

The training approach of the RL Agent is shown in  Figure~\ref{fig:rl}.
%An episode is defined as an entire FL training task which includes $R$ rounds. 
% A step is defined as one round of FL training.
The state is obtained from the Clustering Module and comprises normalized values (for training time and action) in contrast to the observation shown in Figure~\ref{fig2:overview_FedAdapt}.
The RL Agent employs a neural network with three layers that obtains the current input state ($S_t$) as input. 
The RL Agent produces the offloading action $A_t$, which is different to the offloading strategy produced by the Post-processor in that the action is a value between 0 and 1 for a device group, but the offloading strategy is a mapping of this value on to an OP for each device.
The Trained RL Agent is obtained at the end of training. 
The aim of the RL Agent is to maximize the accumulated rewards over each round, which is in line with Equation~\ref{eq:goal1}. The training process begins after the first round with classic FL training (no offloading) \hl{used to} generate the initial state $S_0$.

\subsubsection*{Clustering technique with RL}
A naive design strategy would be to generate an offloading action for each device. However, this strategy would be limited in the following two ways. Firstly, if the number of participating devices changes during FL training, the RL Agent will fail to generate an offloading action due to the fixed input and output dimensions of the neural network used by the RL Agent at the beginning of FL training. Secondly, when the number of devices $K$ becomes large, it is challenging to train the RL Agent due to the large action space that will need to be explored. 
The action space grows exponentially with the increase in the number of devices (for example, consider $K$ devices and a DNN model with $N$ layers, then the size of the action space is $N^K$). 

Therefore, a clustering technique is utilized at the beginning of each FL round (except $S_0$). In the clustering process, homogeneous devices are firstly grouped according to the training time per iteration and network bandwidth between the device and server into $G$ groups (the no. of groups is determined by heuristic algorithms, such as the elbow method~\cite{kodinariya2013review}). Then $G$ groups are used instead of $K$ devices for the input state and output actions dimension.
%The objectives formulated in Equation~\ref{eq:goal1group} is therefore considered:
\di{Therefore, the objective is formulated as:}
%instead of Equation~\ref{eq:goal1} and Equation~\ref{eq:goal2}, respectively.  
\begin{equation}
\di{\begin{aligned}
& \underset{\mu_t^g}{\text{minimize}}
& & \frac{1}{G} \sum_{g=1}^{G} T_t^{g} \\
& \text{subject to}
& & T_t^{g} = \frac{\mu_t^gW^g}{C_t^g} + \frac{(1-\mu_t^g)W^g}{C_t^s} + \frac{L(\mu_t^g)}{Net_t^g}
\label{eq:goal1group}
\end{aligned}}
\end{equation}
where $g$ is defined as a representative device in the group that has the maximum training time. In other words, for each group, \FedAdapt treats all devices in a group as homogeneous devices which means that they have similar computation capability and network bandwidth. In \FedAdapt, the device with the maximum length of training time is used to represent the group. Therefore, $W^g$, $Net_t^g$, $C_t^g$, $\mu_t^g$ and $T_t^{g}$ are bounded by the representative device in each group.
%\begin{equation}
%\begin{aligned}
%& \underset{\mu_t}{\text{minimize}}
%& & \frac{1}{R} \sum_{t=1}^{R} T_t \\
%& \text{subject to}
%& & T_t = \max \{T_t^{g}\}_{g=1}^G \\
%& & & T_t^{k} = f(W^g,C_t^s,C_t^g,Net_t^g,\mu_t^g) \\
%& & & \mu_t = \{\mu_t^g\}_{g=1}^G.
%\label{eq:goal2group}
%\end{aligned}
%end{equation}

\subsubsection*{Optimizing for network bandwidth}
The offloading strategy will need to change when the operational condition, namely network bandwidth between the device and the server changes (the change in OP when network bandwidth changes will be demonstrated in Section~\ref{sec:evaluation}). This is to optimize the performance of \FedAdapt. The RL method will need to  generate a different output action when the network bandwidth changes. 
An intuitive approach is to train the RL Agent for different network bandwidths. However, in practice, we observed that this provides sub-optimal offloading actions due to the rewards that are dominated by when the network bandwidth is not limited. 
To circumvent this, in \FedAdapt, devices with limited network bandwidth are considered within an additional heterogeneous group and devices are dynamically added to this group. At the beginning of each FL round, the network bandwidths of all devices are observed. If the network bandwidth drops below a threshold (discussed in Section~\ref{sec:evaluation}) for a device, then it is assigned to the additional group. The training of the RL Agent is carried out in a controlled environment such that the network bandwidth between the device and the server is limited to represent the group.

\subsubsection*{State and action}
The maximum local training time of the device in a group is used in the input state. The RL Agent in each training round will produce the offloading action for each group. The action of a group is mapped by the Post-processor to the DNNs of all devices in the group, which is $\mu_t^g$. 
For instance, a VGG-5~\cite{simonyan2014very} model with three convolutional layers and two fully connected layers has five offloading actions. 
The output action for each group is designed to be a real value ($\mu_t^g$) ranging from zero to one so that the RL Agent adapts to multiple DNN models. This is mapped to the percentage of the total computational workload of the DNN that is placed on the device. After obtaining $\mu_t^g$, the number of Floating Point Operations (FLOPs) is calculated and set as the target workload on devices. The OP closest to the target workload is chosen. Equation~\ref{eq:in&out} shows input state and output action at round $t$.
\begin{equation}
\begin{aligned}
& S_t = 
& & \{T_t^{g}, \mu_{t-1}^g\}_{g=1}^G \\
& A_t =
& & \{\mu_t^g\}_{g=1}^G \\
& \text{subject to}
& & \mu_t^g \in (0,1]\\
\label{eq:in&out}
\end{aligned}
\end{equation}

\subsubsection*{Reward function}
The reward function guides the training process of the RL Agent. 
The reward obtained at the end of each FL training round is denoted as $R_t$. 
To achieve the objective of Equation~\ref{eq:goal1}, one option is to set the reward as the average training time. 
However, in practice, the device with the largest training time will dominate the reward.
Hence, a normalization function ($f_{norm}$) is used to calculate the reward.
The training time for each device when no DNN model is offloaded is denoted as $B^k$ (a baseline). The training time of device $k$ ($T_t^k$) is normalized with $B^k$ using Equation~\ref{eq:reward}. 
\begin{equation}
\begin{aligned}
& R_t = 
& & \sum_{k=1}^{K} f_{norm}(T_t^{k},{B^k}) \\
& f_{norm} =
& &
    \begin{cases} 
      1 - \frac{T_t^{k}}{B^k} & T_t^{k}\leq B^k \\
      \frac{B^k}{T_t^{k}} - 1 & T_t^{k} > B^k \\
   \end{cases}
\label{eq:reward}
\end{aligned}
\end{equation}

\subsubsection*{Choice of algorithm}
A variety of algorithms are available to train the RL Agent for achieving the objectives presented in Section~\ref{subsec:formulation}; examples include DQN~\cite{sutton1999policy,mnih2013playing} and REINFORCE~\cite{osband2016deep}. In this research, the Proximal Policy Optimization (PPO)~\cite{schulman2017proximal} is chosen. %\footnote{\url{https://openai.com/blog/openai-baselines-ppo/}}
It is a state-of-the-art method which is relatively easy to use and has good performance on standard RL benchmarks~\cite{schulman2017proximal}. Furthermore, compared to DQN that determines the optimal action by evaluating all the possible actions using the Q-network~\cite{mnih2013playing}, PPO generates the output as an explicit action by using a policy network~\cite{schulman2017proximal}. If DQN is adopted for determining the offloading action for each group during FL training, all possible OPs for each group will need to be evaluated using the Q-network~\cite{mnih2013playing}. However, this is not possible in \FedAdapt since the action space is continuous ($\mu_t^g \in (0,1]$) thereby making it impossible to enumerate all actions. Compared to on-policy algorithms, such as REINFORCE, PPO is an off-policy RL algorithm, which repeatedly uses the trajectory data from previous explorations (interactions between the agent and the environment). This improves the training efficiency given that exploration is time consuming. Thus, we choose PPO for the RL algorithm.

\subsubsection*{RL training methodology}
\label{subsubsec:rltrainingmethod}
The RL Agent comprises two fully connected networks, namely the actor and critic networks. Both networks have the same architecture comprising three layers. When the RL Agent is trained the critic network is adopted for assisting the training of the actor network. The actor network is trained to output the offloading action. After completing training, only the actor network will be used to provide the offloading action.
Ideally, the RL Agent should be trained online during an FL task. However, if the RL Agent is trained online during an FL task, the learning time is the time for all rounds in FL training. The RL Agent will need to wait until the completion of each round to obtain the training time required for calculating the reward. 
\di{Therefore, we train the RL Agent in an offline manner before FL tasks. To accelerate the training of the RL Agent, the number of batches used for each round is reduced, referred to as truncated FL rounds in Figure~\ref{fig:rl}. In addition, we collect training time per batch for each device instead of the training time of a round as an element of the input state and output action. The FL model will be trained again with normal rounds (beyond the truncated rounds) after the Trained RL Agent is obtained.}

%\di{\sout{To accelerate the training of the RL Agent, the number of iterations required for each round of FL training is reduced, referred to as truncated FL rounds in Figure~\ref{fig:rl}.The input state and output action are calculated using the training time per iteration for each device instead of the training time of one round for each device. The reduced iterations in each FL rounds will affect the accuracy of the FL model. Since the FL model will be trained with normal rounds (beyond the truncated rounds) after the Trained RL Agent is obtained, the accuracy of the FL model is not considered during the training of the RL Agent.}}

\section{Experimental Studies}
\label{sec:evaluation}
In this section, the performance of \FedAdapt is experimentally verified. The section is organised in response to the research questions raised in Section~\ref{sec:introduction}. Section~\ref{subsec:expart1} presents the results obtained from examining DNN layer offloading in FL (addresses RQ1 on accelerating FL training). 
Section~\ref{subsec:expart2} highlights the results of the RL technique used in \FedAdapt (addresses RQ2 on minimizing the impact of computational heterogeneity of devices). 
Section~\ref{subsec:expart3} presents the results when the RL technique is optimized to account for changing network bandwidth (addresses RQ3). 
The benefits of \FedAdapt are demonstrated by comparing with classic FL. 

\subsection{Layer Offloading in FL}
\label{subsec:expart1}

The assumption that layer offloading can accelerate FL training on computationally limited IoT devices (for example, single board computers, such as Raspberry Pi) is verified. %\footnote{\url{https://www.raspberrypi.org}})
An empirical study is carried out under different network bandwidth values using two Convolutional Neural Networks (CNNs), namely VGG-5~\cite{simonyan2014very} and VGG-8~\cite{simonyan2014very}.
The network bandwidth values correspond to WiFi (75Mbps and 50Mbps; same uplink and downlink bandwidth), 4G+ (25Mbps uplink and 50Mbps downlink) and 4G (10Mbps uplink and 20Mbps downlink) connections.
The study examines the performance of FL for all Offloading Points (OPs) of the CNNs with different network bandwidth.

The study will demonstrate that: (i) Layer offloading from a device to a server reduces the FL training time compared to classic FL in which all layers of the DNN execute on the device. 
Previous studies highlight that computational time on devices is a major bottleneck for resource constrained devices in FL~\cite{he2020group,wang2020towards,gao2020end}.
(ii) The performance gain (reduction in training time) is substantial and offsets the communication overhead that is incurred in transferring the activation and gradient feature maps between the device and the server.

\begin{table}[t] 
\di{\caption{\di{Architecture of the models used for evaluating \FedAdapt. Convolution layers are denoted by C followed by the number of filters; filter size is $3 \times 3$ for all convolution layers, MaxPooling layer is MP, Fully Connected layer is FC with a given number of neurons, and Offloading Point is OP with index.}}
	
	\centering
	\begin{tabular}{ |p{2cm}|p{5cm}| }
    \hline
        Model       &Architecture\\
    
     \hline
        VGG-5        &C32-MP(OP1)-C64-MP(OP2)-C64(OP3)-FC128-FC10(OP4)\\
     \hline
        VGG-8        &C32-C32-MP(OP1)-C64-C64-MP(OP2)-C128-C128(OP3)-FC128-FC10(OP4)  \\
     \hline
    \end{tabular}
	\label{table:vggarcchitectures}}
\end{table}

\subsubsection*{Setup}
The testbed includes an edge server with a 2.5GHz dual-core Intel i7 CPU and an IoT device, namely a Raspberry Pi 4 Model B with 1.5GHz quad-core ARM Cortex-A72 CPU. Only a single device is used to validate the performance gain of FL training with layer offloading. Wi-Fi for the device is supported by the Virgin Media Super Hub 3 router. 

The Linux built-in network traffic control module tc is employed to emulate different bandwidths between the device and the server. %\footnote{\url{https://www.linux.org/docs/man8/tc-netem.html}}
The standard representation of VGG-5 and VGG-8 models used in this study are shown in Table~\ref{table:vggarcchitectures}. For simplicity the batch normalization and non-linear layer (ReLU) are not shown. 
%The size of feature map and kernel are shown as $Channel$ $Number@Height \times Width$ and $Height \times Width$, respectively. 
And the layers denoted with OP present the offloading points empirically tested in this study (all layers after the OP can be offloaded to the server). The CIFAR-10 dataset is used and a batch size of 100 is used for all experiments. 

\subsubsection*{Results}
Table~\ref{table:vgg5res} and Table~\ref{table:vgg8res} \hl{present} the training time per iteration of FL when using layer offloading for all OPs of VGG-5 and VGG-8.
% shown in Figure~\ref{fig:vgg5archi} and Figure~\ref{fig:vgg8archi}, respectively}. 
VGG-5 and VGG-8 both have four OPs (since the FLOPs of the last dense layer is small, an OP is not considered between two dense layers). The last OP in each model (OP4) \hl{corresponds} to device native execution of the DNN as in classic FL. The results are an average of five independent runs. The best result for each value of network bandwidth is in bold. 

\begin{table}[t]
	\caption{Training time per iteration when layer offloading is used in FL for VGG-5 under different network bandwidth.} 
	
	\centering
	\begin{tabular}{ |p{2.5cm}|p{1cm}|p{1cm}|p{1cm}|p{1cm}| }
    %\hline
    %\multicolumn{5}{|c|}{VGG-5 model offloading performance} \\
    \hline
    OP   &75Mbps (Wi-Fi)    &50Mbps  &25Mbps (4G+)    &10Mbps (4G)\\
    
     \hline
     OP1        &\textbf{2.38}   &\textbf{2.7}  &\textbf{3.52}   &6.07\\
     \hline
     OP2        &3.61            &3.9          &4.36   &5.31\\
     \hline
     OP3    &5.24            &5.26         &5.42   &6.73\\
    \hline
     OP4 (device native)          &4.36            &4.36         &4.36   &\textbf{4.36}\\
     \hline
    \end{tabular}
	\label{table:vgg5res}
\end{table}

\begin{table}[t]
	\caption{Training time per iteration when layer offloading is used in FL for VGG-8 under different network bandwidth.}
	
	\centering
	\begin{tabular}{ |p{2.5cm}|p{1cm}|p{1cm}|p{1cm}|p{1cm}| }
    %\hline
    %\multicolumn{5}{|c|}{VGG-8 model offloading performance} \\
    \hline
    OP   &75Mbps (Wi-Fi)    &50Mbps  &25Mbps (4G+)    &10Mbps (4G)\\
     \hline
     OP1        &\textbf{4.75}   &\textbf{5.29}  &\textbf{6.08}   &\textbf{8.84}\\
     \hline
     OP2        &7.52            &8.37           &8.32            &9.95\\
     \hline
     OP3    &10.74           &11.98          &12              &15.93\\
     \hline
      OP4 (device native)          &10.61           &10.61          &10.61           &10.61\\
     \hline
    \end{tabular}
	\label{table:vgg8res}
\end{table}

The best values of training time per iteration for VGG-5 and VGG-8 are $2.38$s and $4.75$s compared to $4.36$s and $10.61$s for classic FL in a Wi-Fi network. The best OP is OP1 both for VGG-5 and VGG-8. Performance is improved when the majority of layers are offloaded from the resource constrained device onto the server. The training time is thus reduced by over 45\% for VGG-5 and over 55\% for VGG-8 compared to classic FL. 

The best OP for both models is the pooling layer -- all layers beyond the pooling layer are offloaded. This is because pooling is a relatively low computationally intensive workload and has a low volume of data (activation feature maps) that needs to be transferred between the device and the server. 

\begin{figure*}[htp]
\begin{center}
	\subfloat[VGG-5]
	{\label{fig:resvgg5}
	\includegraphics[width=0.45\textwidth]
	{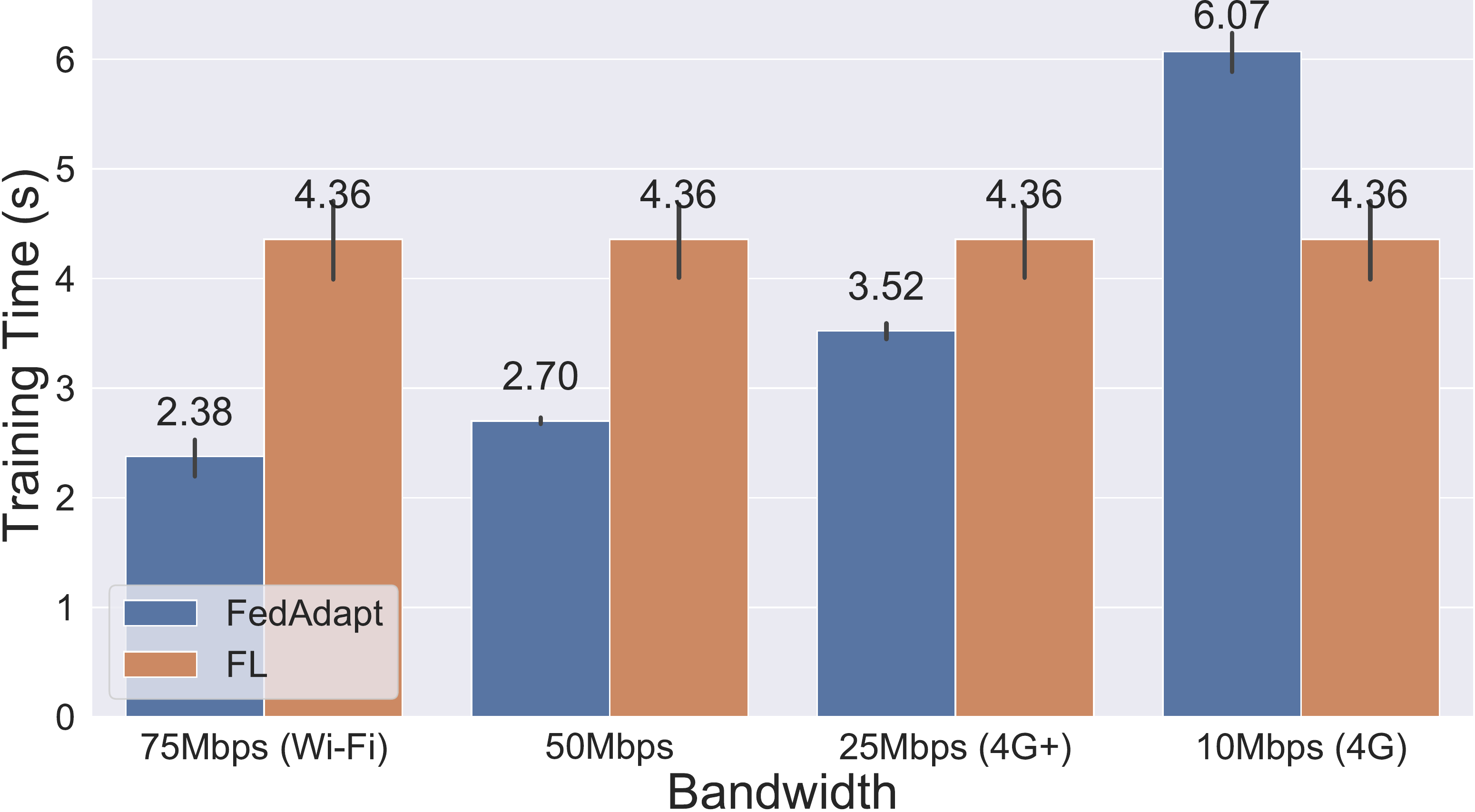}}
	\hfill
	\subfloat[VGG-8]
	{\label{fig:resvgg8}
	\includegraphics[width=0.45\textwidth]
	{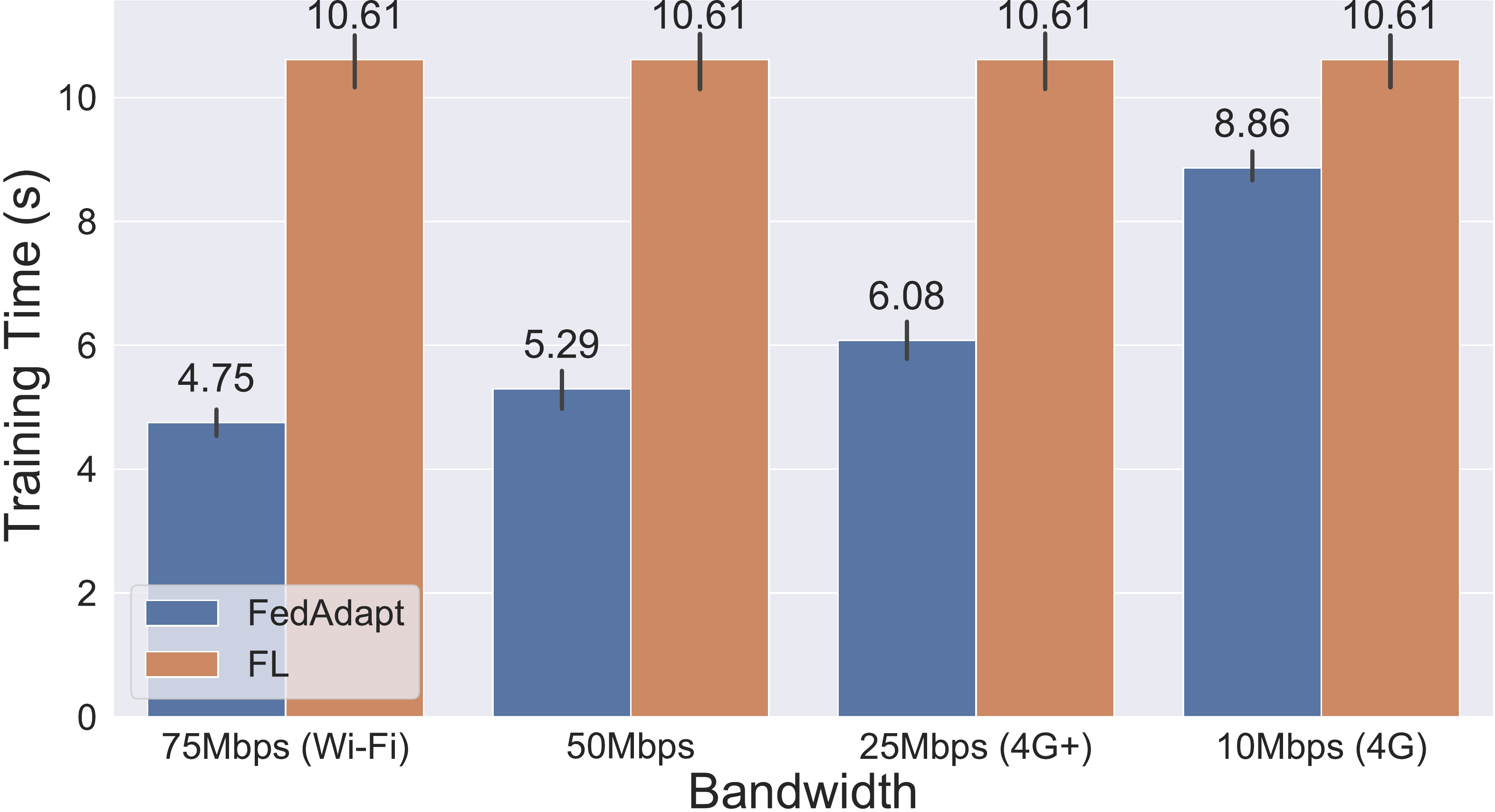}}
\end{center}
\caption{Comparing training time per iteration in \FedAdapt and classic FL}
\label{fig:resvgg}
\end{figure*}

The best performance achieved for the four different network bandwidths is reported in Figure~\ref{fig:resvgg5} and Figure~\ref{fig:resvgg8} for the VGG-5 and VGG-8 model respectively. More than 45\% and 55\% of the training time per iteration can be reduced for the Wi-Fi connection on the VGG-5 and VGG-8, respectively. With the decrease of network bandwidth, the performance acceleration is reduced. For 25Mbps network bandwidth, which is the typical bandwidth of real time 4G+ mobile network\footnote{\url{https://www.4g.co.uk/how-fast-is-4g/}}, the training time is reduced by 19\% and 43\% on VGG-5 and VGG-8, respectively. When the bandwidth is lowered to 10Mbps, offloading has negative effect on the VGG-5 model and the best OP is the OP4 (device native training). However, for the VGG-8 model, offloading reduces 17\% of the training time at 10Mbps bandwidth.

Offloading is influenced by the network bandwidth between the device and the server since there is frequent communication during training. In the case of VGG-5, device native execution is performance efficient for a bandwidth of 10Mbps. However, offloading the layers after OP1 from the device to the server is more effective for higher bandwidths. In the case of VGG-8, offloading is more effective than device native execution. 
Therefore, \FedAdapt is envisioned to be beneficial in scenarios where distributed IoT devices have limited computational resources and need to offload the FL training workload onto a server. Examples include home security cameras that use Wi-Fi and leverage computational resources on a home hub~\cite{gao2020end,nguyen2019diot,musaddiq2018survey} and wearable visual auxiliary equipment that operates in both indoor and outdoor environments (Wi-Fi and on mobile networks)~\cite{chen2020fedhealth,guo2011image,zhuang2020performance}. When the network changes as a device moves across the coverage offered by different networks, then \FedAdapt appropriately selects between device native and offloading-based FL.

\subsection{RL Optimization for Heterogeneity}
\label{subsec:expart2}

Computational heterogeneity of devices leads to the challenge of a straggler (a computationally weak device prolonging the training time of each round) in FL. This is because resource availability of an IoT device will vary due to varying hardware architectures and given that other applications running on the device may require more resources. A straggler device can negatively impact the overall training time. 
Therefore, an offloading strategy that can account for device heterogeneity and minimize the impact of the straggler is required. 

An RL Agent that can select different OPs for the devices in FL is designed. The training process of the RL Agent is guided by the reward function shown in Equation~\ref{eq:reward}. In this experiment, the focus is on device heterogeneity and therefore changing network bandwidth is not considered. 

\subsubsection*{Setup}

The IoT-edge server environment considered has one server and five devices.
The server is the same as used in Section~\ref{subsec:expart1}.
The devices are: (i) Two Raspberry Pi 4 (denoted as Pi$4^1$ and Pi$4^2$) presented in Section~\ref{subsec:expart1}, (ii) Two Raspberry Pi 3 (denoted as Pi$3^1$ and Pi$3^2$) Model B with 1.2GHz quad-core ARM Cortex-A53 CPU, (iii) one Jetson Xavier NX (denoted as Jetson) with embedded GPUs. The running CPU frequency of Pi$4^2$ is set to 0.7GHz to create a straggler in a controlled manner. 

All Raspberry Pis have the same version of Raspbian GNU/Linux 10 (Buster) operating system, Python version 3.7 and PyTorch version 1.4.0. The Jetson and the server have the same version of Python and PyTorch. CuDNN library is installed on the Jetson in order to use the GPU during training. All devices are connected to the server in a Wi-Fi network using a router (presented in Section~\ref{subsec:expart1}). The average available bandwidth between the device and server is 75Mbps. \di{The experiments are carried out in a real-world environment with 5 IoT devices, which is similar to testbeds employed in peer reviewed research on FL~\cite{gao2020end,zhang2021federated,wang2020towards}.}

The DNN model used is VGG-5. It will be demonstrated in Section~\ref{subsubsec:reusingagent} that the RL Agent trained for VGG-5 can also be employed for VGG-8. 
CIFAR-10~\cite{krizhevsky2009learning} is used as the training and testing dataset that contains 50K training and 10K testing samples. The training samples are uniformly divided for the 5 devices without overlapping samples. The entire test dataset is available on the server. The number of FL rounds is 100 and the standard FedAvg~\cite{mcmahan2017communication} aggregation method is used in the server. The horizontal flip technique is used for data augmentation with probability of 0.5 and the Stochastic Gradient Descent (SGD) is utilized as the optimizer for updating the model parameters. The learning rate is $0.01$ at the start of the FL task and $0.001$ at the start of the $50^{th}$ round.

\subsubsection*{Training the RL Agent}

In the experiments reported in this section, the RL agent is first trained and then deployed as a trained agent to the FL task for generating offloading strategies during each round. 
The number of iterations in one round of FL is reduced from 100 to 5 iterations when the RL Agent is trained as presented in Section~\ref{sec:fedadapt-rl}.
The RL Agent has the same training schedule of $50$ rounds. 
PPO is used as the RL algorithm. 
The RL Agent has an actor network and a critic network. 
Both networks have the same architecture of fully connected layers with two hidden layers (64 and 32 neurons, \hl{respectively}). The actor network is used to generate the offloading actions whereas the critic network is used by the RL algorithm to evaluate the value of a given state. During training, a discount factor $\gamma = 0.9$ is set for the RL Agent to determine the importance of using reward from future states and the learning rates for the actor and critic networks are configured to be $1\mathrm{e}{-4}$ (these are standard values used in RL training). The actor and critic networks are updated every $10$ rounds, and during each update the data collected in the previous 10 rounds are used $50$ times. The standard deviation of the actor network is set as $0.5$ at the beginning of the RL training and exponentially decayed (decay rate $0.9$) after $200$ rounds of training. This is to ensure that in the first $200$ rounds, the RL agent has more freedom to explore the action space, but is then decreased to ensure that the RL agent will produce actions that can generate offloading strategies that will reduce the training time. These hyper-parameters remain the same throughout the experiments for better generalization.

\subsubsection*{Clustering}
The initial state $S_0$ is executed without any offloading.
Table~\ref{table:clustering} shows the results of clustering the five devices in the testbed. The Jetson has the faster training speed due to GPU acceleration. Pi$4^2$ is the straggler due to the lower CPU frequency (0.7GHz). Using the values of the training time per iteration of each device the k-means clustering algorithm is used to divide all devices into $G$ groups based on the results from the first round of training. The device training time in the first round is used to cluster the devices into groups. It is assumed that the training speed of devices will not change substantially in subsequent rounds. In this experiment, $G=3$. The Jetson and Pi$4^2$ (0.7GHz) are individually allocated to a group, whereas the Pi$4^1$, Pi$3^1$ and Pi$3^2$ are clustered into one group.
The RL Agent will generate the offloading actions for each group in each round.

\begin{table}[t]
	\caption{Clustering devices into groups when using VGG-5.} 
	
	\centering
	\begin{tabular}{ |p{2cm}|p{1cm}|p{1cm}|p{1cm}| }
    %\hline
    %\multicolumn{4}{|c|}{Clustering result after initial round} \\
    \hline
    Devices   &Training time (s)    &Group no.  &Group center\\
    
     \hline
     Jetson        &0.07   &\textbf{1}  &0.07\\
     \hline
     Pi$4^1$ 1.5GHz        &3.58           &\textbf{2}          &3.7\\
     \hline
     Pi$3^1$ 1.2GHz    &3.75            &\textbf{2}         &3.7\\
    \hline
     Pi$3^1$ 1.2GHz          &3.77            &\textbf{2}        &3.7  \\
     \hline
     Pi$4^2$ 0.7GHz          &5.14            &\textbf{3}         &5.14 \\
     \hline
    \end{tabular}
	\label{table:clustering}
\end{table}

\begin{table}[t]
	\caption{Training time per iteration in seconds for each device for all possible OPs in VGG-5.} 
	
	\centering
	\begin{tabular}{ |p{2.5cm}|p{0.8cm}|p{0.8cm}|p{1.5cm}|p{0.8cm}| }
    %\hline
    %\multicolumn{5}{|c|}{\di{Training time per iteration of each device with different OPs (s)}} \\
    \hline
    OP   &Jetson    &Pi$4^1$ 1.5GHz  &Pi$3^1$ and Pi$3^2$  1.2GHz   &Pi$4^2$ 0.7GHz\\
    
     \hline
     OP1        &0.51   &\textbf{2.38}  &\textbf{2.99}   &\textbf{2.63}\\
     \hline
     OP2        &0.28            &3.61          &3.97   &4.68\\
     \hline
     OP3    &0.27            &5.24         &4.93   &5.88\\
    \hline
     OP4 (device native)          &\textbf{0.17}            &4.36         &4.47   &5.15\\
     \hline
    \end{tabular}
	\label{table:res2devices}
\end{table}

\subsubsection*{Results}
First, all potential OPs for each device are empirically tested to generate the ground truth so as to \hl{verify} the offloading actions produced by the RL Agent and the eventual offloading strategies generated by \FedAdapt; this is shown in Table~\ref{table:res2devices} for VGG-5. The best performance result is shown in bold. All results are an average of five independent runs. 
The optimal offloading actions for each group is $[\mu_{G=1} > 0.96, \mu_{G=2} < 0.38, \mu_{G=3} < 0.38]$. The boundary of all OPs was tested and the borderline between OP1 and OP2 is determined as $0.38$, OP2 and OP3 as $0.79$ and OP3 and OP4 as $0.96$. The proportion of workload (FLOPs) on the device is $0.1$, $0.66$, $0.94$ and $1$ based on the OP. The OP closest to the action generated by the RL Agent is chosen. The boundaries of an OP is the mean of pairwise adjacent OPs ($0.38$, $0.79$, $0.96$). 
Best performance is obtained for all Raspberry Pis when the layers after OP1 are offloaded to the server and for the Jetson is executed device native.

\begin{figure}[t]
		\centering
		\includegraphics[width=0.45\textwidth]{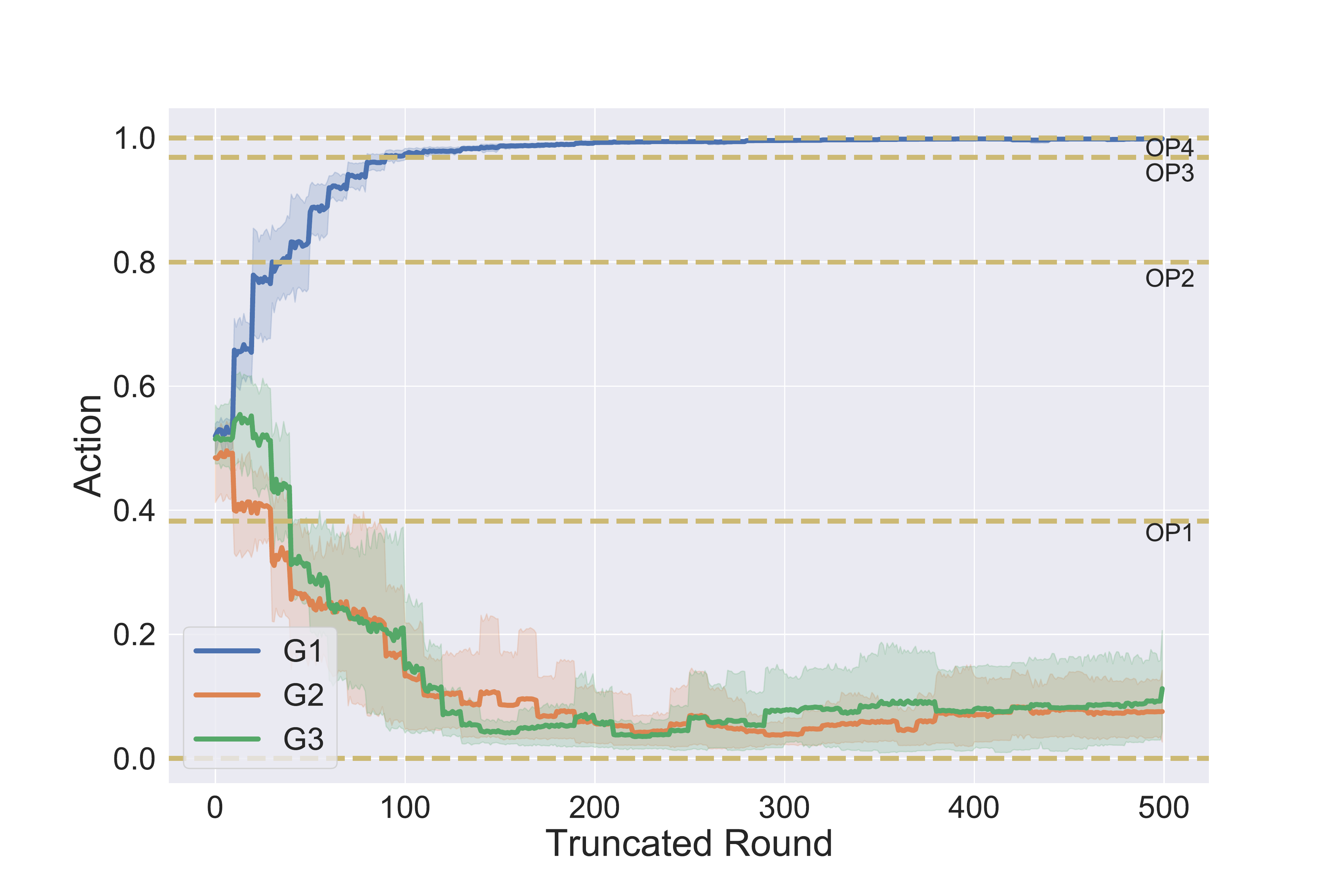}
		\caption{Actions produced for each group chosen by the RL Agent during training for VGG-5.}
		\label{fig:exp2action}
\end{figure}

The empirical results from Table~\ref{table:res2devices} are used to verify the offloading actions of the RL Agent for VGG-5. The action of the RL Agent for 500 rounds (or 500 truncated FL rounds) for VGG-5 is shown in Figure~\ref{fig:exp2action}. The results are the average of five independent runs with different random seeds and are shown for three different groups, $G_1$, $G_2$ and $G_3$. The horizontal lines for OP1, OP2, OP3, and OP4 show the boundaries for each OP.
At the start of RL training, the RL Agent produces similar offloading actions (around $0.5$) for each group. However, the RL Agent optimizes the offloading actions for each group guided by rewards. The mean actions of $G_1$, $G_2$ and $G_3$ become optimal after the $80^{th}$, $30^{th}$, and $40^{th}$ rounds, respectively. After the $80^{th}$ round, the mean actions of all three group are in line with the optimal offloading actions.

\begin{figure}
		\centering
		\includegraphics[width=0.45\textwidth]{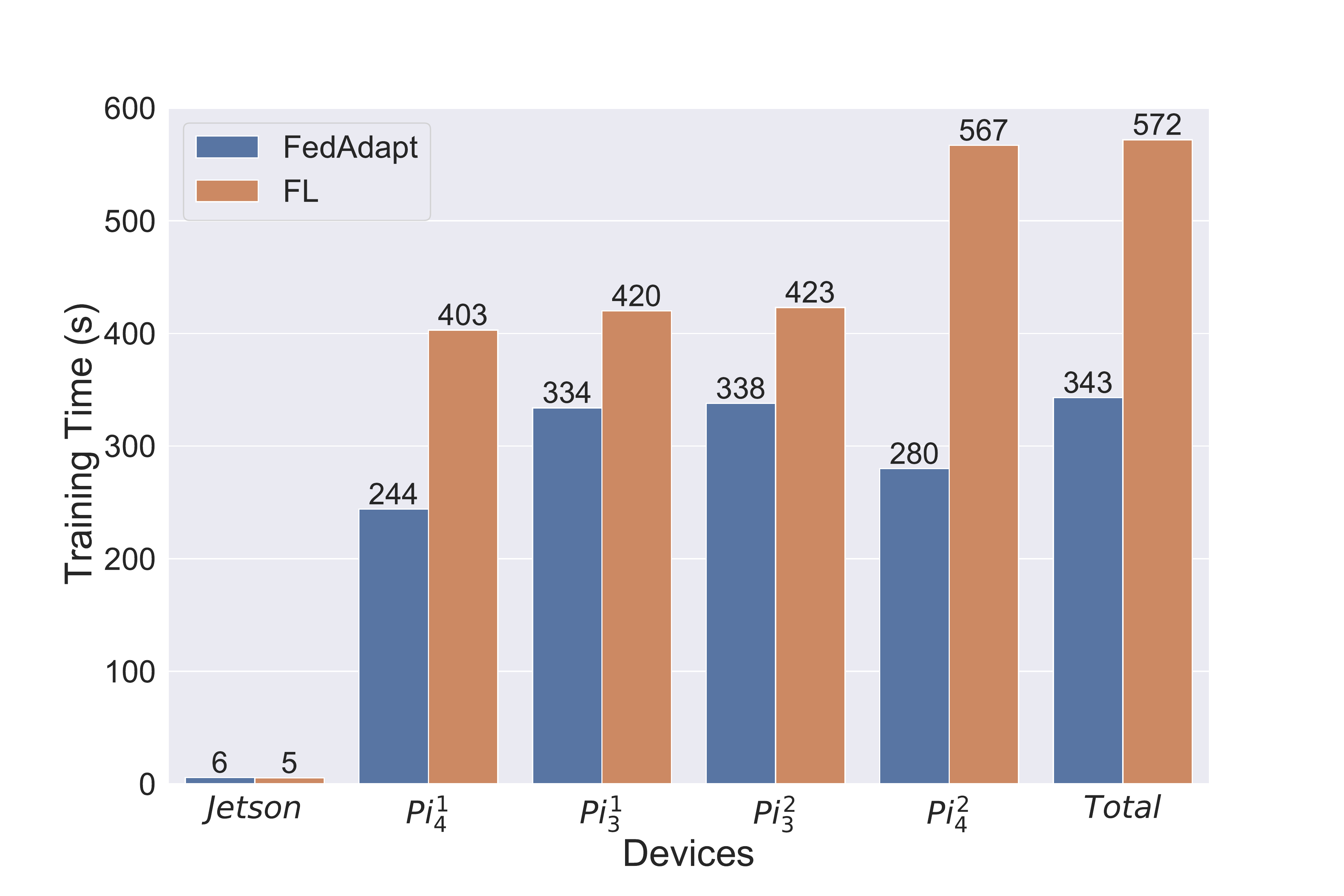}
		\caption{Device and total training time per round in seconds in \FedAdapt and classic FL using VGG-5.}
		\label{fig:exp2devices}
\end{figure}

When the training of the RL agent is complete, the trained actor network is deployed to guide the offloading strategies. The average training time for one round for each device using VGG-5 is shown in Figure~\ref{fig:exp2devices}. The training times for all Raspberry Pis are reduced and any negative impact of offloading is minimized on the Jetson. The maximum performance gain is for the straggler Pi$4^2$ -- a 50\% reduction in training time per round is observed. \FedAdapt saves 40\% of the total training time for one round compared to classic FL. In this experiment, the network bandwidth is not changed during the FL rounds. 

\subsection{Adapting to Changing Network Bandwidth}
\label{subsec:expart3}

This section evaluates \FedAdapt to address RQ3. Limited network bandwidth negatively impacts performance when offloading is used (Table~\ref{table:vgg5res}. For example, if the bandwidth of the device decreases from 75Mbps to 10Mbps then the OP will need to change. 
The experimental setup is similar to Section~\ref{subsec:expart2}.

\subsubsection*{Clustering}
\FedAdapt employs an additional group for devices with low network bandwidth and dynamically groups devices after each round. This is similar to the clustering process presented in Section~\ref{subsec:expart2}, but considers the network bandwidth in addition to the training time per iteration. By considering devices with low bandwidth as a separate group any negative impact on training time is reduced.  
The upload bandwidth of Pi$3^2$ is manually set to 10Mbps (other devices are connected via the 75Mbps Wi-Fi network). Three groups are employed ($G = 3$). The Jetson, Pi$4^1$, Pi$4^2$ and Pi$3^1$ are clustered into two groups based on their training time. The clustering into groups is shown in Table~\ref{table:clustering2}.

\begin{table}[!t]
	\caption{Example of clustering into groups for five devices with one low bandwidth device when using VGG-5.} 
	
	\centering
	\begin{tabular}{ |p{2cm}|p{1cm}|p{1.1cm}|p{1.1cm}|p{1.2cm}| }
    %\hline
    %\multicolumn{5}{|c|}{Clustering results of all devices} \\
    \hline
    Device   &Training time (s)   &Bandwidth  &Group no.  &Group center (s)\\
    
     \hline
     Jetson        &0.07    &75Mbps   &\textbf{1}  &0.07\\
     \hline
     Pi$4^1$ 1.5GHz       &3.58  &75Mbps           &\textbf{2}          &4.16\\
     \hline
     Pi$3^1$ 1.2GHz    &3.75    &75Mbps        &\textbf{2}         &4.16\\
    \hline
     Pi$3^2$ 1.2GHz          &3.77   &10Mbps         &\textbf{3}        &3.7  \\
     \hline
     Pi$4^2$ 0.7GHz          &5.14   &75Mbps       &\textbf{2}         &4.16 \\
     \hline
    \end{tabular}
	\label{table:clustering2}
\end{table}

\subsubsection*{Results}
\begin{figure}[t]
		\centering
		\includegraphics[width=0.48\textwidth]{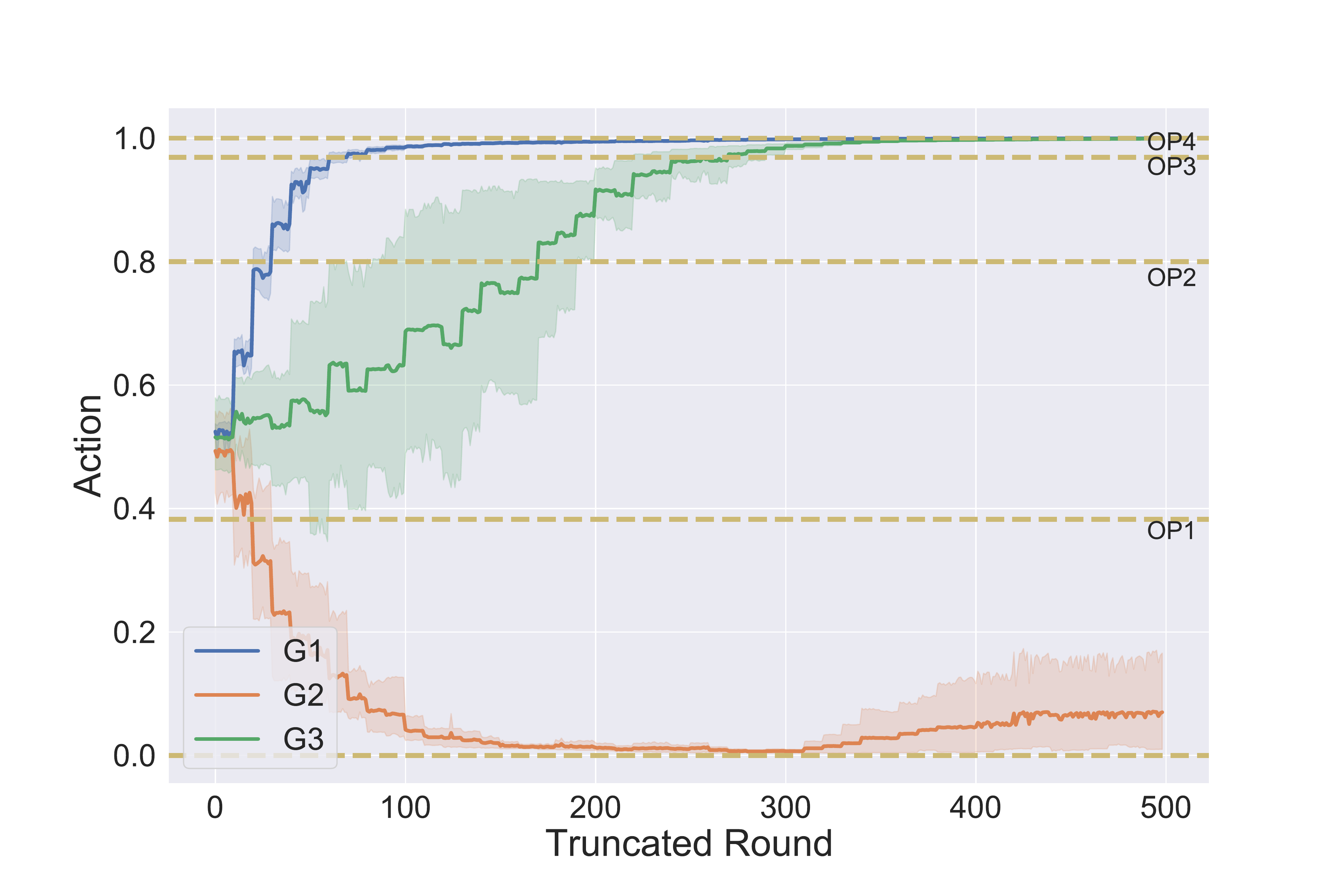}
		\caption{Actions produced for each group chosen by the RL Agent during training of VGG-5 that accounts for devices with low network bandwidth to the server.}
		\label{fig:exp3action}
\end{figure}

The action for each group ($G_1$, $G_2$ and $G_3$) for 500 rounds generated by the RL Agent is shown in Figure~\ref{fig:exp3action} when training VGG-5. The results are the average of five independent runs with different random seeds. The horizontal lines for OP1, OP2, OP3, and OP4 show the boundaries for each OP. 
At the beginning of training, the RL Agent rapidly learns for $G_1$ and $G_2$. After the $20^{th}$ and $60^{th}$ rounds, the mean action of $G_1$ and $G_2$ become optimal, respectively. The optimal action for $G_3$ is determined by the RL Agent only after the $240^{th}$ round with more exploration. This is because at the beginning of training, the reward from $G_1$ and $G_2$ dominate the total reward, which guides the agent in optimizing the action. However, when offloading actions for $G_1$ and $G_2$ become optimal, the agent gradually learns for $G_3$. The mean actions of all three groups are in line with the optimal offloading actions after the $240^{th}$ round.

\subsection{Comparing \FedAdapt and Classic FL}
The performance of \FedAdapt is compared with classic FL on the dimensions of training time and accuracy. The environment is set up on the five devices used previously for FL training on the CIFAR-10 dataset in 100 rounds. During the first 50 rounds of FL training all devices are connected with Wi-Fi (75Mbps bandwidth). The remaining 50 rounds are divided into 5 equal time slots to lower the network bandwidth of the specific devices to 10Mbps in a controlled manner. The sequence is Jetson ($50^{th}$ to $59^{th}$ round), Pi$4^1$ ($60^{th}$ to $69^{th}$ round), Pi$4^2$ ($70^{th}$ to $79^{th}$ round), Pi$3^1$ ($80^{th}$ to $89^{th}$ round), and Pi$3^2$ ($90^{th}$ to $99^{th}$ round).
The \FedAdapt Trained RL Agent from Section~\ref{subsec:expart3} is deployed to produce the offloading action for each device in the FL rounds using VGG-5. 
For classic FL, training on devices is done without offloading.

\begin{figure}[t]
		\centering
		\includegraphics[width=0.48\textwidth]{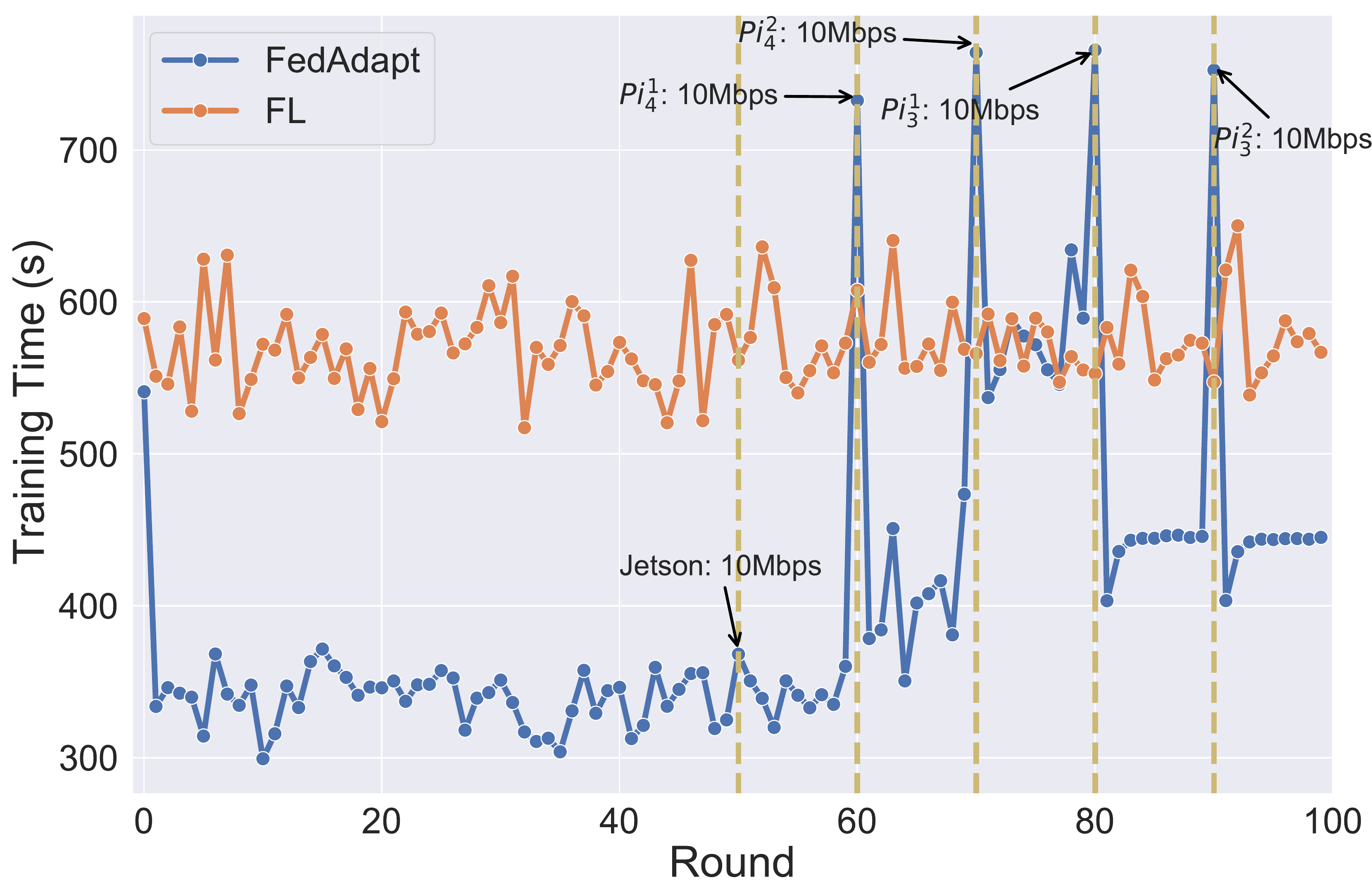}
		\caption{Comparing the training time per round in \FedAdapt and classic FL for VGG-5. Vertical lines show five time slots after the $50^{th}$ round. In each slot, the highlighted device is limited to under 10Mbps bandwidth.}
		\label{fig:exp3time}
\end{figure}

\begin{figure}[ht]
		\centering
		\includegraphics[width=0.48\textwidth]{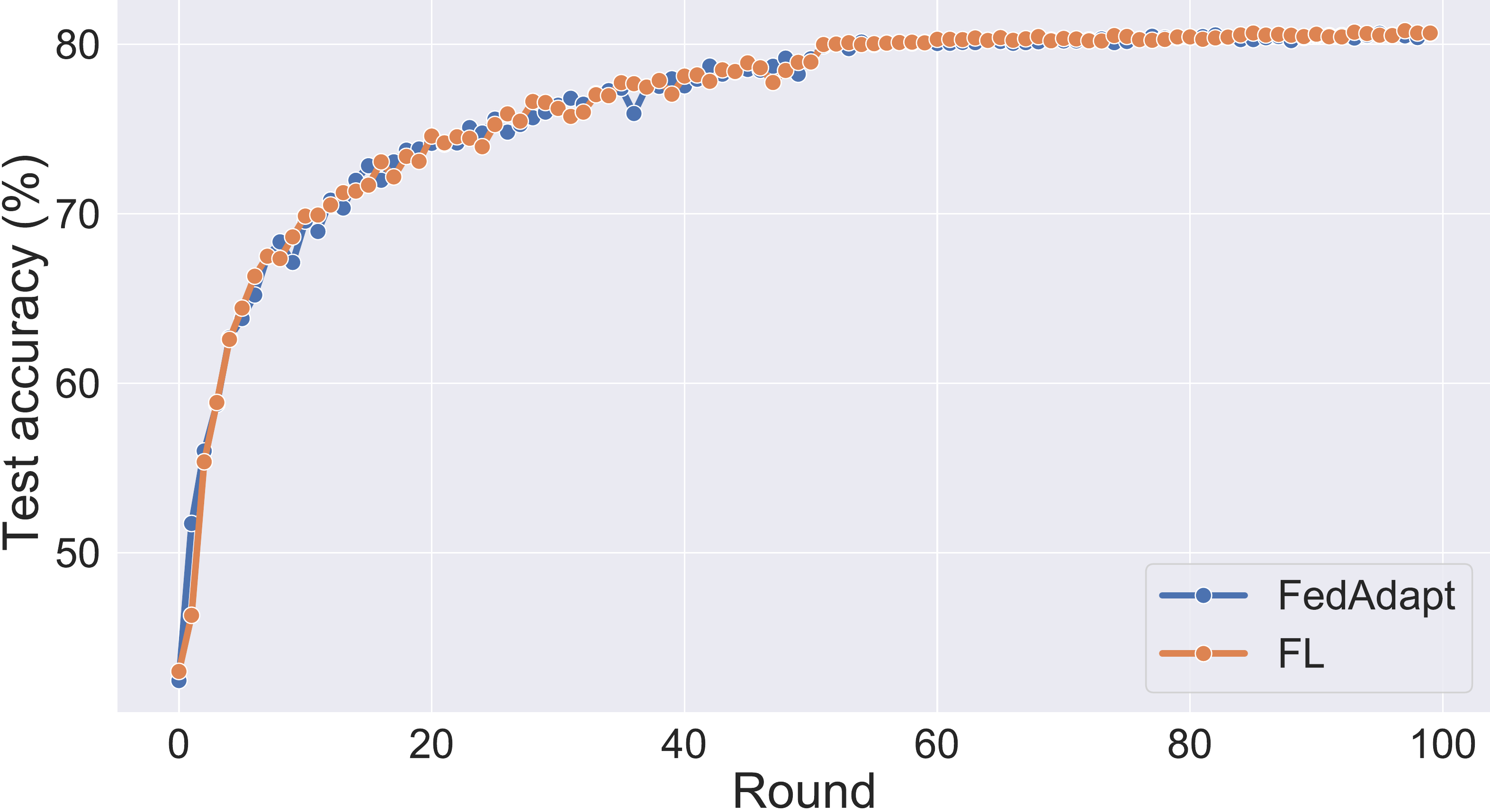}
		\caption{Test accuracy per round of \FedAdapt and classic FL for VGG-5.}
		\label{fig:exp3acc}
\end{figure}

Figure~\ref{fig:exp3time} shows the training time of VGG-5 for each round of \FedAdapt and classic FL. Until the $50^{th}$ round \FedAdapt reduces the average training time by 40\% in comparison to classic FL.  
For the training of the last 50 rounds, the network bandwidth changes for different devices. \FedAdapt responds to the changes by using observations from the previous round. Then a suitable offloading strategy for the current round is obtained. 
Since the optimal action for the Jetson is OP4 (device native), there is limited impact on training time (Round $50$). For the other devices the change of bandwidth makes the offloading action invalid. However, \FedAdapt adapts to these changes in the next round by reassigning the device into $G_3$. For the overall 100 rounds, \FedAdapt reduces the training time by nearly $30\%$ compared to classic FL. 

Figure~\ref{fig:exp3acc} compares the test accuracy of VGG-5 using \FedAdapt and classic FL for 100 rounds. Both have similar accuracy. \FedAdapt employs the FedAvg algorithm as in classic FL. Therefore, \FedAdapt achieves the same convergence speed and final accuracy as classic FL. 
The overhead incurred by \FedAdapt was measured, which comprises the time for running the RL Agent's actor network and the time for redeploying models on each device. An average overhead of $1.6$s was incurred (0.5\% of the time for one round of training). 
In short, the overhead in using \FedAdapt is negligible and the performance gain achieved outperforms classic FL.

\subsubsection*{Reusing the RL Agent of \FedAdapt trained for VGG-5 on VGG-8}
\label{subsubsec:reusingagent}

\begin{figure}
		\centering
		\includegraphics[width=0.45\textwidth]{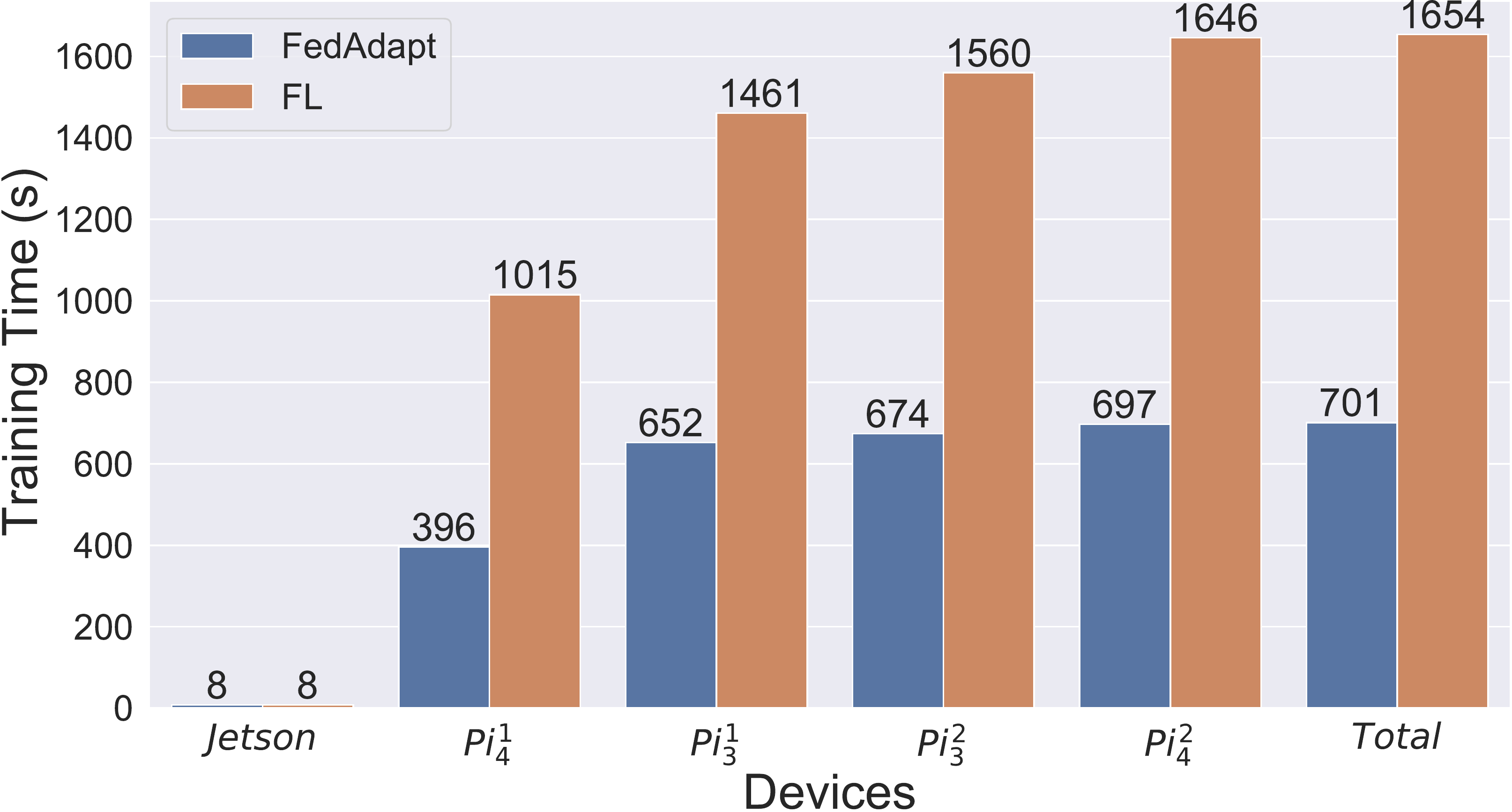}
		\caption{Device and total training time per round in seconds in \FedAdapt and classic FL for VGG-8 when using the RL Agent trained for VGG-5.}
		\label{fig:exp2devicesvgg8}
\end{figure}

\begin{figure}
		\centering
		\includegraphics[width=0.48\textwidth]{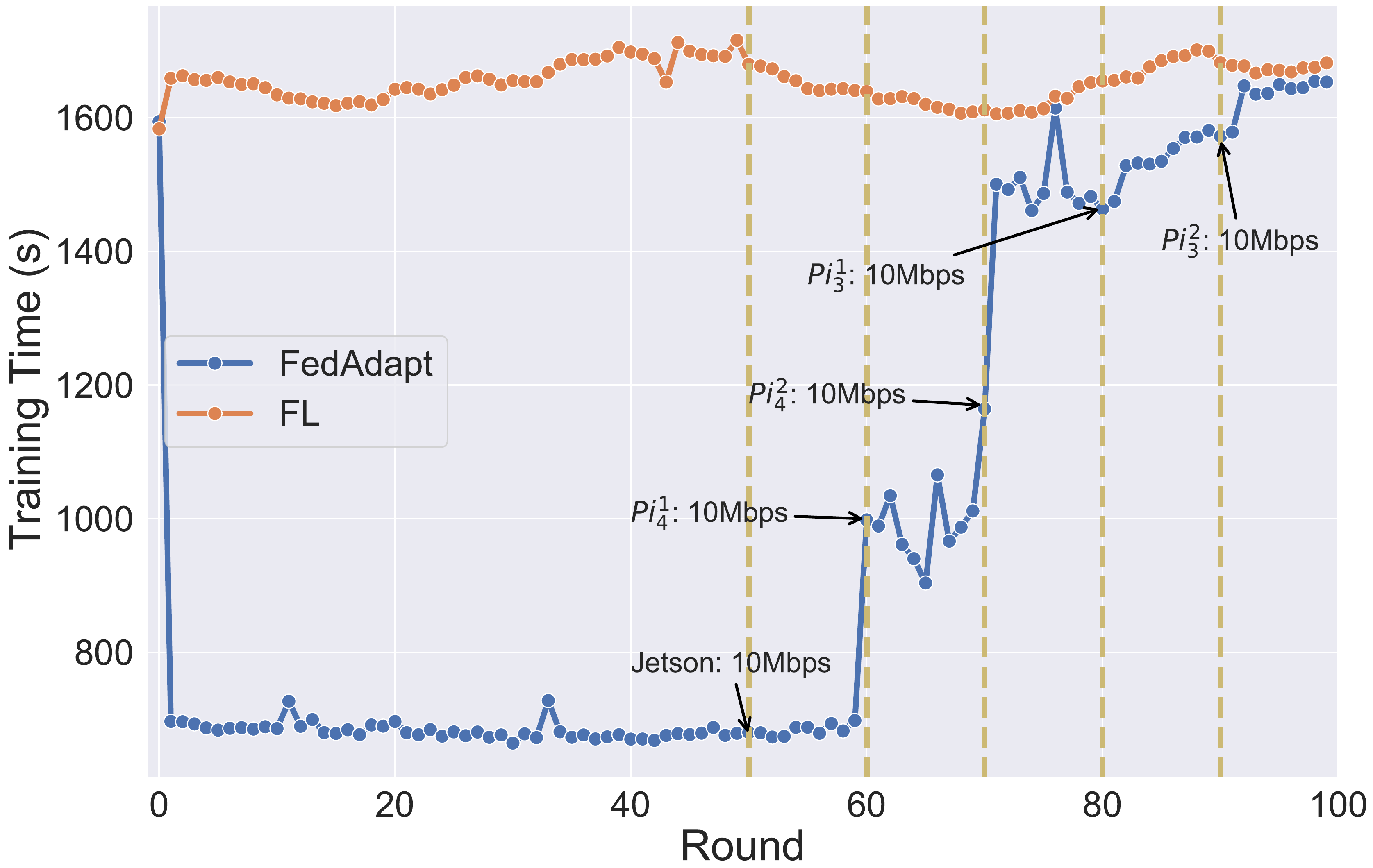}
		\caption{Comparing the training time per round in \FedAdapt and classic FL for VGG-8 when using the RL Agent trained for VGG-5. %Vertical lines show five time slots after the $50^{th}$ round. In each slot, the highlighted device is limited to under 10Mbps bandwidth.
		Experimental setting is presented in Section~\ref{subsec:expart3}.
		}
		\label{fig:exp3timevgg8}
\end{figure}

The performance of \FedAdapt using the RL Agent that was trained for VGG-5 was evaluated on VGG-8.
The Trained RL Agent on VGG-5 (presented in Section~\ref{subsec:expart3}) is used without custom retraining for VGG-8. The average training time of one round for each device is shown in Figure~\ref{fig:exp2devicesvgg8}. The maximum performance gain is for the straggler Pi$4^2$ -- a 57\% reduction in training time per round is observed. Overall, \FedAdapt saves 57\% of training time compared to classic FL. 
Although \FedAdapt reduces the training time when the network bandwidth changes, it generates sub-optimal offloading strategies after the $70^{th}$ round. 
Instead of selecting offloading-based strategies for devices that have a low bandwidth for VGG-8, the RL Agent selects device native strategies, thereby minimizing the performance gain. 
This is because the RL Agent is trained for VGG-5 in which device native strategies have maximum performance gain. 
Nonetheless, the overall training time is reduced by nearly $40\%$ when compared to classic FL as shown in Figure~\ref{fig:exp3timevgg8}.

\section{Conclusions}
\label{sec:conclustion}
%Classic Federated Learning (FL) is impractical in IoT-edge environments for three reasons. Firstly, computationally constrained devices have large training times. Secondly, computational heterogeneity of devices result in stragglers that will slow down training. Finally, varying operational conditions, such as network bandwidth between the device and server, can significantly affect training performance.

\di{
Classic FL is impractical in IoT-edge environments given the limited computational capacity on IoT devices, heterogeneity of devices and varying network bandwidth between the device and server, all of which significantly affect training performance.
%Classic FL is impractical in IoT-edge environments due to resource constrained and heterogeneous devices and varying network bandwidth between the device and server, which significantly affect training performance. 
This paper presented \FedAdapt, a holistic framework that surmounts the above limitations by incorporating three techniques for accelerating FL, reducing the impact of stragglers and adapting to varying network bandwidth.}
\di{\FedAdapt reduces the training time of stragglers by over half compared to classic FL. When faced with stragglers and changing network bandwidth \FedAdapt outperformed classic FL by reducing training time up to 40\% while achieving the same accuracy and convergence speed with negligible execution time overhead.}
%\di{\sout{The first is an offloading technique in which the layers of a DNN model are offloaded from the device to a server for alleviating the computational burden of training on resource constrained devices.
%The second is a RL technique to automatically identify the layers that need to be offloaded from the device to the server for heterogeneous devices to reduce the impact of stragglers. Finally, an optimization technique for RL is employed to account for changing network bandwidth. A clustering technique is implemented to rapidly generate the offloading strategy for all devices.
%The above were developed on a testbed comprising five devices and a server and using the CIFAR-10 dataset.The key results are that \FedAdapt introduces a negligible execution time overhead that is offset by performance gains. \FedAdapt reduces the training time of stragglers by over half the time compared to classic FL. When faced with stragglers and changing network bandwidth \FedAdapt outperformed classic FL by reducing training time up to 40\% while achieving the same accuracy and convergence speed as classic FL.}}

\textit{Limitations and Future Work:} \FedAdapt relies on RL and clustering for generating offloading strategies for each participating device. However, as the number of devices becomes large, a single RL agent may not be suitable to generate offloading strategies for all devices. Distributed RL agents will need to be investigated along with hierarchical clustering of devices. 
In addition, since \FedAdapt employs offloading-based training to accelerate training on IoT devices, additional communication overheads between the devices and the server are introduced. Techniques such as quantization may reduce the communication cost. These will be explored in future work.

\bibliographystyle{IEEEtran}
\bibliography{FedAdapt}

\vskip -2.5\baselineskip plus -1fil
\begin{IEEEbiography}[{\includegraphics[width=1in,height=1.25in,clip,keepaspectratio]{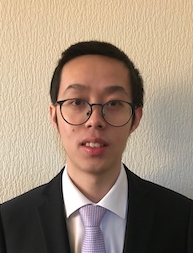}}]{Di Wu} is currently pursuing a PhD degree in computer science at University of St Andrews, UK. He received a B.S. degree in Information System and Information Management from Northeast Forestry University, China in 2015, and an M.S. degree in Data Science from University of Southampton, UK in 2018. His major interests are in the areas of federated learning, distributed machine learning, edge computing, model compression, and Internet-of-Things.
\end{IEEEbiography}
\vskip -2.5\baselineskip plus -1fil

\begin{IEEEbiography}[{\includegraphics[width=1in,height=1.35in,clip,keepaspectratio]{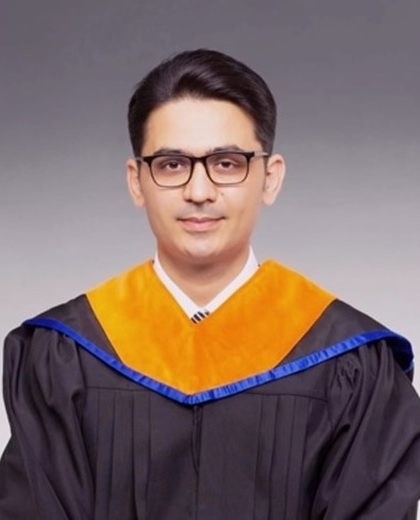}}]{Rehmat Ullah} earned a PhD degree in electronics and computer engineering from Hongik University, South Korea. He is currently a research fellow at the University of St Andrews, UK. Previously, he worked as a research fellow at Queen's University Belfast, UK and as an assistant professor at Gachon University, South Korea. His research interests are in edge computing, information centric networking and 5G evolution and beyond with a recent focus on federated learning for edge computing systems.  More information is available from \url{www.rehmatkhan.com}.
\end{IEEEbiography}
\vskip -2.5\baselineskip plus -1fil

\begin{IEEEbiography}[{\includegraphics[width=1in,height=1.25in,clip,keepaspectratio]{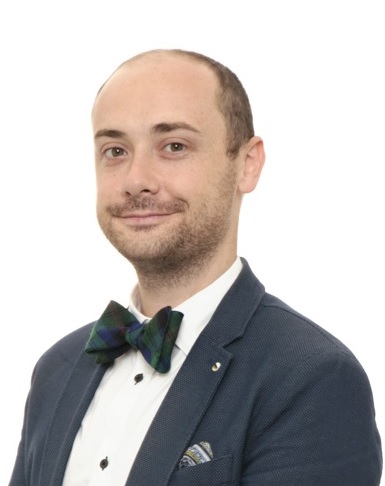}}]{Paul Harvey} received the PhD in computing science from the University of Glasgow, UK. He is one of the original founders of the Autonomous Networks Research and Innovation Lab in Rakuten Mobile, Japan, and is a co-chair in the ITU focus group on autonomous networks. He is a research lead at Rakuten Mobile.
\end{IEEEbiography}
\vskip -2.5\baselineskip plus -1fil

\begin{IEEEbiography}[{\includegraphics[width=1in,height=1.25in,clip,keepaspectratio]{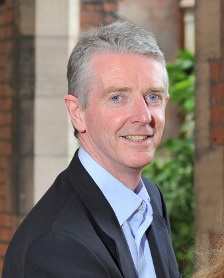}}]{Peter Kilpatrick} is a Reader in computer science at Queen's University Belfast. His interests include parallel programming models and cloud and edge computing. 
\end{IEEEbiography}
\vskip -2.5\baselineskip plus -1fil

\begin{IEEEbiography}[{\includegraphics[width=1in,height=1.25in,clip,keepaspectratio]{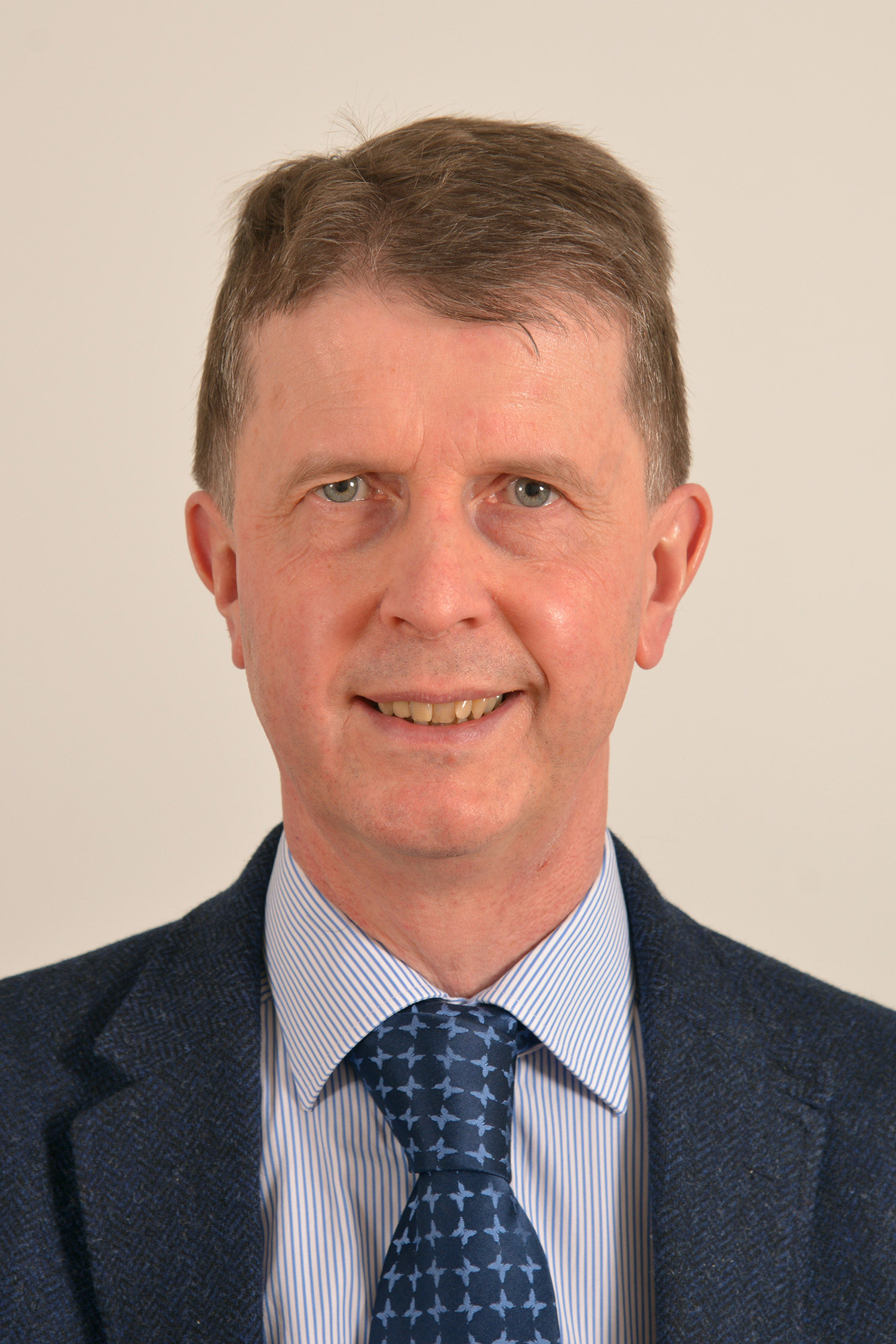}}]{Ivor Spence} received the PhD degree in computer science from Queen's University Belfast, UK, where he did research on code generation. He is currently a Reader in computer science at Queen's University Belfast where he leads the artificial intelligence (AI) research theme.
His research is primarily on heterogeneous computing systems for AI.
\end{IEEEbiography}
\vskip -2.5\baselineskip plus -1fil

\begin{IEEEbiography}[{\includegraphics[width=1in,height=1.25in,clip,keepaspectratio]{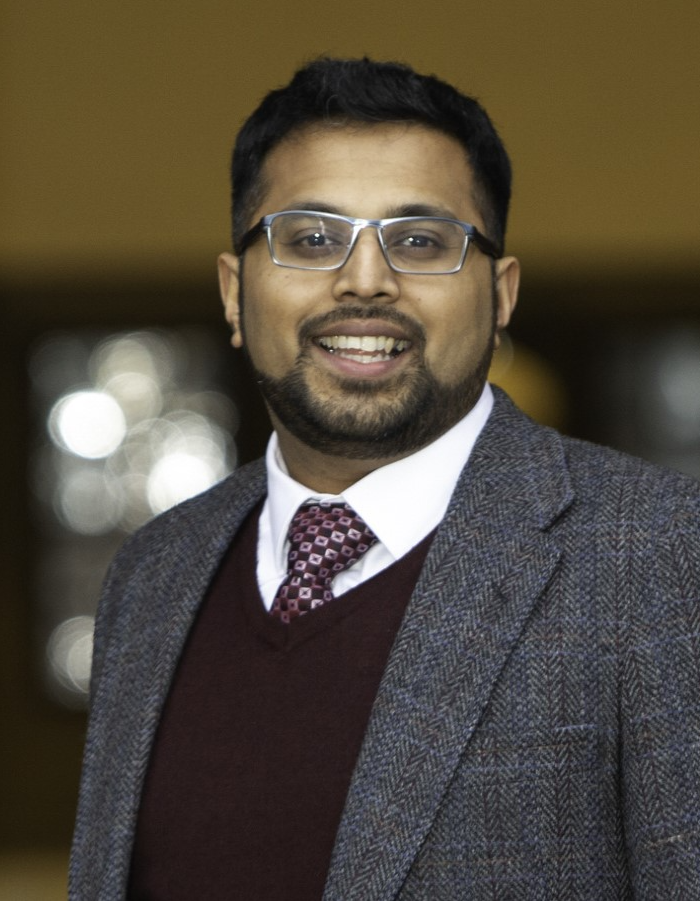}}]{Blesson Varghese} received the PhD degree in computer science from the University of Reading, UK on international scholarships. He is a Reader in computer science at the University of St Andrews, UK, and the Principal Investigator of the Edge Computing Hub. He is an Honorary faculty member at Queen's University Belfast and a previous Royal Society Short Industry Fellow. His interests include distributed systems that span the cloud-edge-device continuum and edge intelligence applications. More information is available from \url{www.blessonv.com}.
\end{IEEEbiography}
\newpage
\end{document}